\documentclass[epj]{svjour}
\usepackage{graphicx,figsize}

\begin{document}
\title{Non-extensive Statistics and Understanding Particle Production and Kinetic Freeze-out Process from $p_T$-spectra at 2.76 TeV}

\author{Bhaskar De\thanks{e-mail: bhaskar.de@gmail.com}}

\institute{Department of Physics,
A. P. C. Roy Government College,
Himachal Bihar, Matigara, Siliguri-734010,
West Bengal, India.
}

\date{}

\abstract{
An approach, based on Tsallis non-extensive statistics, has been employed, here, to analyse, systematically, the $p_T$-spectra of various identified secondary hadrons like pions, kaons, protons and antiprotons, produced in different central $Pb+Pb$ interactions at LHC energy 2.76 TeV in terms of multiplicity and temperature fluctuations. The results, thus obtained, have been utilized to understand the various stages of different types of hadron production during evolution of the fireball produced in such collisions.
 % end of PACS codes
}

\authorrunning{B. De}
\titlerunning{Non-extensive Statistics and Understanding Particle Production and Freeze-out Process at 2.76 TeV}
\maketitle

\newpage
%\doublespacing
\section{Introduction}
The heavy ion collisions at ultra-relativistic energies are supposed to provide significant clues on the possible occurrence of a phase transition from a confined hadronic state to a deconfined plasma state made of elementary constituents of hadrons. However, no direct information on the properties of such a hot and dense partonic matter, if formed in such high energy nuclear interactions, can be obtained as it does not live long enough to extract any direct information on its equation of state. It readily expands, cools down and undergoes further phase transition which results in emission of thousands of different variety of secondary hadrons. So, these hadron-spectra are the only, though indirect, sources to have an understanding, in detail, of the course of it's excursion from formation to hadronization.
\par
The transverse momentum spectra of hadrons are treated as one of the important tools to understand the dynamics of high energy collisions. The systematic analysis with the help of an appropriate model or approach of such an observable may throw light on various thermodynamical as well as hydrodynamical properties of the fireball at different stages of its evolution. In a very recent study, the present author had taken up such an effort in dealing with pion- and kaon-spectra produced in $P+P$ and $Pb+Pb$ collisions at $\sqrt{s_{NN}}=2.76$ TeV\cite{De1}. The main theoretical basis of the analysis was Tsallis non-extensive statistics which has, so far, been proven to be very useful in interpreting various aspects of high energy nuclear interactions by different theoretical as well as experimental groups\cite{Tsallis1,Tsallis2,Tsallis3,Tsallis4,Tsallis5,Prato1,Beck1,Beck2,Beck3,Wilk1,Wilk1.1,Wilk1.2,Wilk1.3,Wilk1.4,Wilk2,Wilk3,Wilk4,Wilk5,Osada1,Biro1,Biro2,Biro3,Biro3.1,Biro4,Urmossy0.1,Biyajima1,Biyajima2,Alberico1,Lavagno1,Lavagno2,Kaniadakis,Kodama,De2,De3,Wibig1,Wibig2,Jiulin1,Jiulin2,Urmossy1,Urmossy2,Deppman1,Deppman2,Deppman3,Deppman4,Deppman5,Cleymans1,Cleymans2,Cleymans3,Cleymans4,Cleymans5,Rybczynski1,Rybczynski2,Ristea1,Khandai1}.       
\par
In the present study, we have, once again, taken up such a systematic analysis for the same set of interactions at the same energy. However, the present effort differs from the previous one in two ways: (i) The earlier version was confined only to two lightest mesonic varieties($\pi$- and $K$-mesons) while the periphery of the present study has further been widened by accommodating proton/antiproton-spectra. (ii) One of the essential parameter, the transverse collective flow, was not taken into account in our previous theoretical approach. But, such collective motions in the expanding hot and dense partonic fluid are supposed to contribute more to the average transverse momenta of the heavier hadrons keeping the average thermal momentum same for all the varieties\cite{Lee1,Blaizot1,Bearden1}. So, transverse flow plays a crucial role in determining the spectral shape of the high-mass secondaries like protons. This particular parameter has now been incorporated in our present theoretical approach. 
\par
The organization of the present work is as follows: a brief
outline of the theoretical development of the main working formulae,
to be used in the present study, has been presented in next section. The obtained results and a detailed discussion on it have been provided in Section 3. And the last section is preserved for the concluding remarks.

\section{Outline of the Theoretical Approach}

The generalized statistics of Tsallis is not only applicable to an
equilibrium system, but also to the nonequilibrium systems with
stationary states\cite{Beck2}. As the name `nonextensive' implies,
these entropies are not additive for independent systems.
\par
The nonextensive Boltzmann factor is defined as\cite{Beck2}
\begin{equation}
x_{ij} ~ = ~ (1 ~ + ~ (q-1)\beta \epsilon_{ij})^{-q/(q-1)}
\end{equation}

 where $\epsilon_{ij} ~ = ~ \sqrt{\textbf{p}_i^2 ~ + ~ m_j^2}$ is the energy associated with the $j-$th particle of rest mass $m_j$ in the momentum state $i$,  $\beta=1/T$ is the inverse temperature variable, $q$ is a measure of degree of fluctuation present in the system and is called nonextensivity parameter; with $q\rightarrow1$, the above equation approaches the ordinary Boltzmann factor $e^{-\beta\epsilon_{ij}}$.
\par
 If $\nu_{ij}$ denotes
the number of particles of type $j$ in momentum state $i$, the
generalized grand canonical partition function is given by,

\begin{equation}
Z ~ = ~ \sum_{(\nu)} \prod_{ij} x_{ij}^{\nu_{ij}}
\end{equation}

The average occupation number of a particle of species $j$ in the
momentum state $i$ can be written as\cite{Beck2}

\begin{equation}
{\bar{\nu}_{ij}}= x_{ij}\frac{\partial}{\partial x_{ij}} \log{Z} ~
= ~ \frac{1}{(1+(q-1)\beta\epsilon_{ij})^{q/(q-1)}\pm 1}
\end{equation}

where $-$ sign is for bosons and the $+$ sign is for fermions.
\par
The probability of observation of a particle of mass $m_0$ in a certain
momentum state can be obtained by multiplying the average occupation number with
the available volume in the momentum space\cite{Beck2}. The infinitesimal volume in momentum space
is given by

\begin{equation}
d^3p ~ = ~ E ~ dy ~ p_T ~ dp_T ~ d\phi
\end{equation}

where $E$ is the energy, $p_T$ is the transverse momentum, $y$ is the rapidity and $\phi$ is the azimuthal angle.\\
If the particle-spectra is influenced by the presence of the hydrodynamical flow in the system, the energy associated with the detected secondary can be written as\cite{Lavagno2},
\begin{equation}
E ~ = ~ v^{\mu}p_{\mu}
\end{equation}

where $v^{\mu} ~ = ~ \gamma(1, ~ \vec{v})$ \\
and $p^{\mu} ~ = ~ (p_0,\vec{p}) ~ = ~ (m_T ~ coshy, ~ \vec{p_T}, ~ m_T ~ sinhy)$ are the hydrodynamic four-velocity and the four-momentum of the particle respectively with $\gamma=\frac{1}{\sqrt{1- |\vec{v}|^2}}$ being the Lorentz factor and $m_T=\sqrt{m_0^2+p_T^2}$ being the transverse mass of the detected secondary.
\par
If we neglect the effect of longitudinal flow over that of the transverse flow of particles in a co-moving system in the central rapidity region($y\simeq 0$), the expressions for four-velocity and four momentum are given by,

\begin{equation}
v^{\mu} ~ = ~ \gamma(1, ~ \vec{v_T}, ~ 0)
\end{equation}
and 
\begin{equation}
p^{\mu} ~ =  ~ (m_T , ~ \vec{p_T}, ~ 0)
\end{equation}
with $\gamma=\frac{1}{\sqrt{1-{v_T}^2}}$.
\par
Further, if we assume that $\vec{v_T}$ and $\vec{p_T}$ are collinear, the energy-term will take the form  
\begin{equation}
E ~ = ~ \gamma (m_T ~ - ~ \vec{v_T}.\vec{p_T}) ~ = ~ \gamma (m_T ~ - ~ v_T p_T) 
\end{equation}
where $v_T$ is the average transverse velocity.
\par
Hence, the probability density $w(p_T,y)$  is given by:

\begin{equation}
\frac{1}{2\pi} ~ \frac{d^2N}{p_Tdp_Tdy}  ~ = ~ C ~ 
\frac{\gamma (m_T ~ - ~ v_T p_T)}{[1+(q-1)\beta \gamma (m_T ~ - ~ v_T p_T)]^{q/(q-1)} \pm 1}
\end{equation}
where $C$ is a proportionality constant.
\par
Using the relationships $\beta = \frac{1}{T_{eff}}$,  where $T_{eff}$ is the effective temperature of the interaction region, the invariant yield at mid-rapidity(for $y\simeq 0$ ) will take the form

\begin{equation}
\frac{1}{2\pi} ~ \frac{d^2N}{p_Tdp_Tdy}  ~ = ~ C ~ 
\frac{\gamma (m_T ~ - ~ v_T p_T)}{[1+(q-1) ~ \frac{\gamma (m_T ~ - ~ v_T p_T)}{T_{eff}}]^{q/(q-1)} \pm 1}
\end{equation}
\par
The average multiplicity of the detected secondary per unit rapidity in the given rapidity region can be obtained by the relationship
\begin{equation}
\begin{array}{lcl}
\frac{dN}{dy} & = &  \int_0^\infty \frac{d^2N}{dp_T ~ dy} ~ dp_T \\
&  & \\ 
& = & C_1 ~
\int_0^\infty \frac{\gamma (m_T ~ - ~ v_T p_T)}{[1+(q-1) ~ \frac{\gamma (m_T ~ - ~ v_T p_T)}{T_{eff}}]^{q/(q-1)} \pm 1} p_T dp_T
\end{array}
\end{equation}
where $C_1=2\pi C$.\\
Hence, the constant $C_1$ can be expressed in terms of $\frac{dN}{dy}$ by the relationship

\begin{equation}
C_1 = \frac{dN}{dy} ~ \frac{1}{\int_0^\infty \frac{\gamma (m_T ~ - ~ v_T p_T)}{[1+(q-1) ~ \frac{\gamma (m_T ~ - ~ v_T p_T)}{T_{eff}}]^{q/(q-1)} \pm 1} p_T dp_T}
\end{equation}

\par
Combination of eqn(10) and eqn(12) will provide us the main working formula for invariant yield for a detected secondary and it is given by
\begin{equation}
\begin{array}{lcl}
\frac{d^2N}{p_T ~ dp_T ~ dy} ~ & = & ~ \frac{dN}{dy} ~ \frac{1}{\int_0^\infty \frac{\gamma (m_T ~ - ~ v_T p_T)}{[1+(q-1) ~ \frac{\gamma (m_T ~ - ~ v_T p_T)}{T_{eff}}]^{q/(q-1)} \pm 1} p_T dp_T} \\
& & \\
&  & \times \frac{\gamma (m_T ~ - ~ v_T p_T)}{[1+(q-1) ~ \frac{\gamma (m_T ~ - ~ v_T p_T)}{T_{eff}}]^{q/(q-1)} \pm 1}
\end{array}
\end{equation}

\par
Further, it was observed earlier that the parameters $T_{eff}$ and $q$ are strongly correlated, even if they were set free\cite{Wilk3,De2}. So, these two parameters alongwith the average multiplicity can phenomenologically be correlated  by the following relationships\cite{De1,Wilk3,De3}:

\begin{equation}
T_{eff}=T_{kin}(1-c(q-1))
\end{equation}

\begin{equation}
\frac{<N> - n_0N_{part}}{<N>}=c(q-1)
\end{equation}

where $T_{kin}$ is the kinetic freeze-out temperature, $<N>$ is the average multiplicity of the detected secondary produced in $A+A$ interactions and $n_0$ is the same for $P+P$ interactions. However, if the studied rapidity regions and their widths($\Delta y$) are same or nearly same for both the cases, one can replace $<N>$ and $n_0$ with the corresponding rapidity yields($\frac{dN}{dy}$). In the present study, the data, under consideration for both the systems, are available from the central rapidity region($|y|<0.5$ with $\Delta y = 1$); and hence, we can set $<N>=\frac{dN}{dy}$. 
\par 
Here, equation (14) takes into account the fluctuation in effective temperature while that in multiplicity by equation(15). These two types of fluctuations are mutually correlated through the factor $c(q-1)$ where $c$ is the parameter which takes care of the fluctuations of the system arising out of a stochastic process in any selected region of the system and/or of some energy transfer between the selected region and the rest of the system\cite{Wilk3}. However, for the sake of calculational simplicity it is assumed that $c$ is independent of any flow-velocity. 
\par
Furthermore, in our previous study\cite{De1}, it was observed that, in some cases, the product term($n_0N_{part}$), in the last equation, exceeds $<N>$ which makes $c$ negative. But, negativity of $c$ violates the assumption that the transfer of energy takes place only from the interaction region to the spectators of the non-interacting nucleons\cite{Wilk3}. Hence, to keep the physical assumptions, associated with the above constraints, valid, the `$-$' sign was replaced with `$\sim$' in our previous work\cite{De1}, so that only the magnitude of the fluctuation between $<N>$ and $n_0N_{part}$ could be taken into account; and the modified form of the second constraint is given by,

\begin{equation}
\frac{<N> \sim n_0N_{part}}{<N>}=c(q-1)
\end{equation}
\par
Equation(13) alongwith equations(14) \& equation(16) forms the bais of our theoretical analysis of the transverse momentum spectra from nucleus-nucleus collisions.
\section{Results and Discussions}
The identified hadron spectra produced in $P+P$ interactions have been fitted with the basic working formula given in eqn.(13) excluding the constraints given in eqn.(14) and eqn.(16). The outcomes are presented in graphical format in Fig.1 and in tabular form in Table-1, where $n_o$ denotes the average multiplicity per unit rapidity of the produced hadron-variety in $P+P$ interaction. It is quite clear from Table-1 that the effect of transverse flow on particle spectra is quite insignificant as far as $P+P$ interactions are concerned as the number of participant nucleons and hence the interaction volume is quite low in these cases.
\par
Fig2-Fig4 depict the fits to the experimental data on transverse momentum spectra for production of pions, kaons and proton-antiprotons in different central $Pb+Pb$ collisions at 2.76 TeV at LHC. All these fits have, now, been obtained on the basis of
equation(13) alongwith the constraints given in eqn(14) \& eqn.(16). The values of various parameters 
obtained from the fits are provided in Table-2-Table-4. All the fits can be treated as quite satisfactory on the basis of obtained values of $\chi^2/ndf$ given in Table-1-Table-4. 
\par
Four figures in Fig.5 represent the behaviour of four parameters which carry very important information on the dynamical properties of the fireballs produced in different central $Pb+Pb$ interactions. 
\par
The centrality-dependence of two parameters, the effective temperature $T_{eff}$ and and the non-extensive parameter $q$ are provided graphically in Fig.5(a) and Fig.5(b) respectively. The desired anti-correlation between these two parameters is, in general, quite visible for all the varieties except few cases for k-meson production in central interactions. 
\par
The kinetic freeze-out temperature, $T_{kin}$, obtained from  $\pi^{0,\pm}$, $K^{\pm}$ and $p/\bar{p}$-spectra from different central $Pb+Pb$ collisions are depicted in Fig.5(c). It is observed from the figure that this particular parameter is coming out to be constant and the average value of it is around 167 MeV for kaons and 236 MeV for protons. But, as far as pi-mesons are concerned, $T_{kin}$ shows an increasing trend from 95 MeV for most central(0-5$\%$) collision to around 160 MeV for the most peripheral(80-90$\%$) one. In our earlier studies\cite{De1,De3} with different mesonic varieties at LHC as well as at RHIC energies this particular parameter was identified as Hagedorn's critical temperature($T_0$) due to it's constant behaviour for all the varieties and it's order of magnitude was around $\sim m_{\pi}$. But, in the present study, where the effect of transverse flow has been taken into account, the deviation in the parameter's behaviour from the earlier findings drives us to interpret it in a different way. 
\par
Three different values of $T_{kin}$ for three different types of secondaries produced in a particular centrality indicate that the kinetic freeze-out temperature is quite different for different secondaries. More massive a secondary hadron is, the corresponding freeze-out temperature is much higher which can only be interpreted as the formations of the heavier seondaries cease at much earlier stages of the evolution compared to lighter varieties.  In this sense, the protons/anti-protons do not form in appreciable number, on the average, below 236 MeV while the temperature around 167 MeV is the thermal boundary for $K$-meson production.   
\par
The number of participant nucleons($N_{part}$) in the overlap area between two colliding nuclei and hence the interaction volume in the fireball decreases from central to peripheral collisions, which results in lesser number of binary collisions arising out of rescattering processes suffered by the constituting partons as well as the hadrons already produced at the initial stages of the evolution of the hot fireball. This certainly delays the overall freeze-out process when one goes from the peripheral to central collisions. The values of $T_{kin}$ extracted from pion-spectra reflect this particular fact as it is quite low for central collisions compared to peripheral ones. So, we can infer that these particular values of the kinetic freeze-out temperature provide the information on the final freeze-out temperatures of the fireballs at respective centralities when all sort of particle interactions come to an end. And pi-meson is the most abundant variety which is produced during the period from time of freeze-out for K-meson production to final freeze-out.         
\par
Fig.5(d) represents the graphical nature of average transverse flow $v_T$ as a function of number of participant nucleons. The nature is some what similar for all the varieties while going from central to peripheral collisions. But, it is quite sensitive to the mass of the detected secondary as far as it's magnitude is concerned, which increases from pi-meson to protons for a particular central collision.  
\par
The values of another parameter $c$ corresponding to different spectra have been given in Table-2-Table-4. It is seen that $c$ has, on average, an increasing tendency for the meson-spectra while going from central to peripheral collisions. This observation is in accordance with our earlier study\cite{De1}. But, for proton-spectra, this very parameter exhibits, though very weak, a decreasing trend.   
\par 
The integrated yield per unit rapidity($<N>=\frac{dN}{dy}$) in the central rapidity region for different varieties produced in $P+P$ and $Pb+Pb$ interactions have been provided in Table-1-Table-4. The results corresponding to mesonic varieties are somewhat similar to our earlier findings\cite{De1}. The overall findings on $<N>$ are, in general, in good agreement with the results reported in Ref.\cite{Abelev4}. The obtained values, on the basis of present analysis, have been represented in three different ways in Fig.6(a)-6(c). Fig.6(a) depicts the results as a function of $N_{part}$ while Fig.6(b-c) represent the same, but this time normalized by per pair of participant nucleons($N_{part}/2$) and by pair of participant quarks($N_{q-part}/2$) respectively. The values of $N_{part}$ for different centralities have been obtained from Ref.\cite{Abelev2,Aamodt1,Abelev3} and those of $N_{q-part}$(Table-5) have been calculated using PHOBOS Glauber Monte Carlo Simulation\cite{Alver1} alongwith with the method suggested in Ref.\cite{Voloshin1} and applied in Ref.\cite{Netrakanti1,De4}. Fig.6(b)-Fig.6(c) indicates a nearly linear dependence of $<N>$ on $N_{q-part}$ instead of on $N_{part}$ for all the varieties including the protons and it's antiparticle. This is, as expected, in accordance with our earlier studies\cite{De1,De4}. 

\section{Conclusions}
In this study we have presented an analysis,in the light of Tsalli's non-extensive statistics, of the identified hadron spectra --- from pions to protons --- produced in $Pb+Pb$ interactions at LHC energy 2.76 TeV. The study can be treated as a sequel to our earlier analysis\cite{De1} after making a very necessary modification in terms of collective transverse flow in the theoretical approach to get more insights on the hydrodynamical evolution of the produced fireball while dealing with transverse momentum spectra, specially for heavier secondaries. The final modified working formula is quite successful in reproducing the experimental data as far as $\chi^2/ndf$ is concerned.   
\par
The dependence of $v_T$ on centralities and on mass of the detected secondaries, found in the present analysis for LHC energy, is quite similar with those found in a recent study\cite{Khandai1} which had dealt with identified hadron-spectra including heavier baryons like $\Lambda$ and $\Xi$ at RHIC energies with the aid of somewhat similar approach sans any predefined correlation between $T_{eff}$ and $q$. But, the present findings on the behaviour of the freeze-out temperature are in sharp contrast with those found for RHIC energy. From the present analysis, $T_{kin}$ is found to be somewhat constant for kaons and protons while for pions, it shows strong centrality dependence. On the otherhand, the corresponding values obtained in Ref.\cite{Khandai1} exhibit constant or rather weak dependence on centralities for pion- and kaon-spectra, whereas for baryon-spectra the freeze-out temperature shows strong centrality-dependence. The deviation in $T_{kin}$, observed in our present analysis, can be attributed to the constraints imposed by eqn(14) \& eqn.(16). However, the present interpretation of freezing out of heavier hadrons at earlier stages finds support in the above-referenced work.
\par
Besides, the ALICE experimental group had analysed the same spectra using a variant of blast-wave model\cite{Abelev4}. Their analytical result on the centrality-dependence of $v_T$ is quite similar with respect to the present analysis, except from some mismatch in magnitude. And the values of final freeze-out temperature are in good agreement with the present findings.
\par
So, in brief, we can conclude, on the basis of the discussions made in the previous two paragraphs, that the present approach can be treated as potentially fruitful to draw valuable insights on hydrodynamical as well as thermodynamical picture of the evolution and freeze-out of the fireball produced due to deposition of enourmous energy in relativistic heavy ion collision, despite some of it's shortcomings like proper identification of the parameter $c$.  However, in the present analysis, we could not deal with other heavy baryons like $\Lambda$ and $\Xi$ as the present approach needs the spectra for a particular variety from both the nucleon-nucleon and nucleus-nucleus interactions at the same energy due to presence of eqn(14) \& eqn.(16). We hope that the present approach would be able to throw more light on the dynamics of high energy collisions if and when the data on spectra over a wide $p_T$-range for other heavier secondaries produced both in $P+P$ and $A+A$ interactions will be available.      
\par
\begin{acknowledgement}
The author is indebted to the Learned Referee for his/her valuable suggestions for the betterment of an earlier version of the present manuscript.
\end{acknowledgement}

\newpage
\begin{table*}[h]
\caption{Values of fitted parameters with respect to the experimental
data on hadron-spectra produced in $P+P$ collision at $\sqrt{s_{NN}}=2.76$ TeV}
\ \\
\centering
\begin{tabular}{lllllll}
\hline \noalign{\smallskip} 
Secondary type & $N_{part}$ & $n_0$ & $q$ & $T_{eff}$(GeV) & $v_T$(c) & $\chi^2/ndf$\\
\hline \noalign{\smallskip}
$\pi^0$ &  & $1.89\pm0.03$  & $1.148\pm0.004$ & $0.082\pm0.002$ & $0.010\pm 0.001$ & $8.594/15$ \\
$\pi^+$ &  & $1.89\pm0.06$  & $1.149\pm0.004$ & $0.080\pm0.003$ & $0.010\pm 0.002$ & $9.746/35$ \\
%\hline
$\pi^-$ & 2 & $1.89\pm0.05$  & $1.150\pm0.004$ & $0.079\pm0.002$ & $0.010\pm 0.002$ & $10.985/35$ \\
%\hline
$K^+$ &  & $0.241\pm0.004$  & $1.148\pm0.003$ & $0.083\pm0.002$ & $0.030\pm 0.001$ & $17.614/35$ \\
%\hline
$K^-$ &  & $0.241\pm0.003$  & $1.148\pm0.004$ & $0.083\pm0.004$ & $0.032\pm 0.002$ & $21.492/35$ \\
%\hline
$p$ &  & $0.103\pm0.002$  & $1.120\pm0.004$ & $0.083\pm0.002$ & $0.010\pm 0.001$ & $15.748/40$ \\
%\hline
$\bar{p}$ &  & $0.104\pm0.003$  & $1.122\pm0.002$ & $0.082\pm0.002$ & $0.010\pm 0.002$ & $10.460/40$ \\
\noalign{\smallskip} \hline \noalign{\smallskip}
\end{tabular}
\end{table*}
\begin{table*}[h]
\caption{Values of fitted parameters with respect to the experimental
data on pion-spectra at different centralities of $Pb+Pb$ collisions at LHC energy $\sqrt{s_{NN}}=2.76$ TeV}
\begin{center}
{\small{
\begin{tabular}{llllllll}
\hline \noalign{\smallskip}
Secondary & Centrality & {    $<N>$  } & {    $v_T(c)$  } & {    $T_{kin}$(GeV)  } & {$ c$   } & $q$ & $\chi^2/ndf$ \\
type &  &  &  &  &  &  &  \\ \hline \noalign{\smallskip}

 {$\pi^0$} & $0-20\%$ & {503$\pm$ 11} & {0.29$\pm$ 0.03} & {0.100$\pm$ 0.002} & {1.343$\pm$  0.004} & 1.117 & 3.920/6 \\ 
 {} & $20-40\%$ & {247$\pm$ 8} & {0.28$\pm$ 0.02} & {0.104$\pm$ 0.002} & {1.677$\pm$ 0.005} & 1.120 & 3.816/6 \\ 
 {} & $40-60\%$ & {103$\pm$ 4} & {0.20$\pm$ 0.01} & {0.111$\pm$ 0.003} & {2.048$\pm$ 0.003} & 1.130 & 2.391/6 \\ 
 {} & $60-80\%$ & {30.7$\pm$ 0.5} & {0.19$\pm$ 0.02} & {0.130$\pm$ 0.002} & {2.992$\pm$ 0.002} & 1.139 & 6.643/6 \\ \hline

 {$\pi^+$} & $0-5\%$ &  {734$\pm$ 22} &  {0.29$\pm$ 0.02} &  {0.096$\pm$ 0.003} &  {0.140$\pm$  0.004} & 1.102 & 39.607/37 \\ 
 {} & $5-10\%$ &  {605$\pm$ 17} &  {0.27$\pm$ 0.02} &  {0.097$\pm$ 0.002} &  {0.282$\pm$ 0.005} & 1.106 & 34.759/37 \\ 
 {} & $10-20\%$ &  {455$\pm$ 14} &  {0.26$\pm$ 0.03} &  {0.098$\pm$ 0.002} &  {0.713$\pm$ 0.004} & 1.115 & 29.530/37 \\ 
 {} & $20-30\%$ &  {305$\pm$ 11} &  {0.25$\pm$ 0.02} &  {0.099$\pm$ 0.002} &  {1.196$\pm$ 0.003} & 1.130 & 27.888/37 \\ 
 {} & $30-40\%$ &  {199$\pm$ 10} &  {0.23$\pm$ 0.02} &  {0.101$\pm$ 0.004} &  {1.617$\pm$ 0.007} & 1.139 & 23.285/37 \\ 
 {} & $40-50\%$ &  {122$\pm$ 7} &  {0.22$\pm$ 0.01} &  {0.105$\pm$ 0.003} &  {2.156$\pm$ 0.005} & 1.147 & 23.080/37 \\ 
 {} & $50-60\%$ &  {70$\pm$ 6} &  {0.21$\pm$ 0.02} &  {0.112$\pm$ 0.004} &  {2.663$\pm$ 0.008} & 1.160 & 22.339/37 \\ 
 {} & $60-70\%$ &  {37$\pm$ 0.8} &  {0.20$\pm$ 0.03} &  {0.128$\pm$ 0.005} &  {3.248$\pm$ 0.006} & 1.164 & 18.912/37 \\ 
 {} & $70-80\%$ &  {19$\pm$ 0.6} &  {0.19$\pm$ 0.03} &  {0.137$\pm$ 0.006} &  {3.543$\pm$ 0.004} & 1.161 & 20.144/37 \\ 
 {} & $80-90\%$ &  {8.3$\pm$ 0.2} &  {0.16$\pm$ 0.03} &  {0.162$\pm$ 0.006} &  {4.064$\pm$ 0.006} & 1.174 & 64.583/37 \\ \hline \noalign{\smallskip}

 {$\pi^{-}$} & $0-5\%$ &  {734 $\pm$23} &  {0.31 $\pm$0.02} &  {0.095 $\pm$0.002} &  {0.143 $\pm$0.004} & 1.100 & 37.544/37 \\ 
 {} & $5-10\%$ &  {605 $\pm$19} &  {0.29 $\pm$0.03} &  {0.096 $\pm$0.002} &  {0.280 $\pm$0.003} & 1.107 & 31.600/37 \\ 
 {} & $10-20\%$ &  {454 $\pm$11} &  {0.27 $\pm$0.02} &  {0.098 $\pm$0.003} &  {0.725 $\pm$0.003} & 1.117 & 26.937/37 \\ 
 {} & $20-30\%$ &  {307 $\pm$12} &  {0.25 $\pm$0.02} &  {0.100 $\pm$0.002} &  {1.204 $\pm$0.004} & 1.123 & 25.828/37 \\ 
 {} & $30-40\%$ &  {200 $\pm$12} &  {0.24 $\pm$0.03} &  {0.102 $\pm$0.002} &  {1.602 $\pm$0.003} & 1.136 & 22.715/37 \\ 
 {} & $40-50\%$ &  {123 $\pm$8} &  {0.22 $\pm$0.03} &  {0.106 $\pm$0.003} &  {2.128 $\pm$0.006} & 1.144 & 18.213/37 \\ 
 {} & $50-60\%$ &  {70 $\pm$7} &  {0.20 $\pm$0.02} &  {0.117 $\pm$0.005} &  {2.734 $\pm$0.004} & 1.156 & 18.217/37 \\ 
 {} & $60-70\%$ &  {37 $\pm$0.4} &  {0.19 $\pm$0.04} &  {0.130 $\pm$0.005} &  {3.263 $\pm$0.004} & 1.163 & 18.197/37 \\ 
 {} & $70-80\%$ &  {19 $\pm$0.5} &  {0.18 $\pm$0.05} &  {0.139 $\pm$0.003} &  {3.566 $\pm$0.004} & 1.160 & 20.886/37 \\ 
 {} & $80-90\%$ &  {8.3 $\pm$0.3} &  {0.18 $\pm$0.03} &  {0.160 $\pm$0.007} &  {4.092 $\pm$0.007} & 1.173 & 64.842/37 \\ \noalign{\smallskip} \hline \noalign{\smallskip}
\end{tabular}
}}

\end{center}
\label{}
\end{table*}

\begin{table*}[p]
\caption{Values of fitted parameters with respect to the experimental
data on kaon-spectra at different centralities of $Pb+Pb$ collisions at LHC energy $\sqrt{s_{NN}}=2.76$ TeV}
\begin{center}
{\small{
\begin{tabular}{llllllll}
\hline \noalign{\smallskip}
Secondary & Centrality & {    $<N>$  } & {    $v_T(c)$  } & {    $T_{kin}$(GeV)  } & {$ c$   } & $q$ & $\chi^2/ndf$ \\
type &  &  &  &  &  &  &  \\ \hline \noalign{\smallskip}
 {$K^{+}$} & $0-5\%$ &  {109 $\pm$ 8} &  {0.41 $\pm$ 0.03} &  {0.166 $\pm$ 0.002} &  {2.840 $\pm$ 0.005} & 1.054 & 3.380/32 \\ 
 {} & $5-10\%$ &  {90 $\pm$ 6} &  {0.36 $\pm$ 0.01} &  {0.167 $\pm$ 0.002} &  {1.967 $\pm$ 0.004} & 1.060 & 4.767/32 \\ 
 {} & $10-20\%$ &  {66 $\pm$ 6} &  {0.32 $\pm$ 0.02} &  {0.166 $\pm$ 0.003} &  {0.866 $\pm$ 0.002} & 1.056 & 5.399/32 \\ 
 {} & $20-30\%$ &  {46 $\pm$ 7} &  {0.3 0$\pm$ 0.02} &  {0.167 $\pm$ 0.003} &  {0.424 $\pm$ 0.001} & 1.055 & 6.611/32 \\ 
 {} & $30-40\%$ &  {31.5 $\pm$ 3.4} &  {0.27 $\pm$ 0.03} &  {0.165 $\pm$ 0.003} &  {0.238 $\pm$ 0.002} & 1.058 & 15.945/32 \\  
 {} & $40-50\%$ &  {18 $\pm$ 4} &  {0.26 $\pm$ 0.02} &  {0.167 $\pm$ 0.004} &  {1.847 $\pm$ 0.006} & 1.075 & 7.151/32 \\ 
 {} & $50-60\%$ &  {10.1 $\pm$ 2.1} &  {0.25 $\pm$ 0.01} &  {0.168 $\pm$ 0.005} &  {2.875 $\pm$ 0.006} & 1.090 & 7.555/32 \\ 
 {} & $60-70\%$ &  {5.1 $\pm$ 1.1} &  {0.24 $\pm$ 0.02} &  {0.168 $\pm$ 0.005} &  {3.612 $\pm$ 0.004} & 1.116 & 7.279/32 \\ 
 {} & $70-80\%$ &  {2.4 $\pm$ 0.6} &  {0.22 $\pm$ 0.03} &  {0.168 $\pm$ 0.006} &  {4.013 $\pm$ 0.008} & 1.146 & 8.953/32 \\ 
 {} & $80-90\%$ &  {0.974 $\pm$ 0.002} &  {0.17 $\pm$ 0.03} &  {0.169 $\pm$ 0.008} &  {4.357 $\pm$ 0.007} & 1.196 & 33.064/32 \\ \hline
 {$K^{-}$} & $0-5\%$ &  {109 $\pm$ 9} &  {0.40 $\pm$ 0.04} &  {0.167 $\pm$ 0.003} &  {3.041 $\pm$ 0.005} & 1.051 & 6.040/32 \\ 
 {} & $5-10\%$ &  {90 $\pm$ 6} &  {0.35 $\pm$ 0.03} &  {0.167 $\pm$ 0.003} &  {1.970 $\pm$ 0.005} & 1.059 & 3.914/32 \\ 
 {} & $10-20\%$ &  {66 $\pm$ 8} &  {0.30 $\pm$ 0.03} &  {0.167 $\pm$ 0.002} &  {0.866 $\pm$ 0.003} & 1.056 & 5.354/32 \\ 
 {} & $20-30\%$ &  {46 $\pm$ 8} &  {0.28 $\pm$ 0.03} &  {0.166 $\pm$ 0.002} &  {0.420 $\pm$ 0.003} & 1.056 & 6.250/32 \\ 
 {} & $30-40\%$ &  {31.9 $\pm$ 2.8} &  {0.25 $\pm$ 0.04} &  {0.165 $\pm$ 0.002} &  {0.238 $\pm$ 0.002} & 1.058 & 17.030/32 \\  
 {} & $40-50\%$ &  {18 $\pm$ 4} &  {0.24 $\pm$ 0.02} &  {0.168 $\pm$ 0.004} &  {1.838 $\pm$ 0.004} & 1.075 & 7.551/32 \\ 
 {} & $50-60\%$ &  {10 $\pm$ 2} &  {0.23 $\pm$ 0.02} &  {0.167 $\pm$ 0.003} &  {2.880 $\pm$ 0.003} & 1.095 & 9.768/32 \\ 
 {} & $60-70\%$ &  {5.1 $\pm$ 0.8} &  {0.24 $\pm$ 0.02} &  {0.167 $\pm$ 0.004} &  {3.614 $\pm$ 0.005} & 1.116 & 7.269/32 \\ 
 {} & $70-80\%$ &  {2.3 $\pm$ 0.2} &  {0.21 $\pm$ 0.02} &  {0.168 $\pm$ 0.007} &  {4.016 $\pm$ 0.004} & 1.163 & 7.563/32 \\ 
 {} & $80-90\%$ &  {0.974 $\pm$ 0.003} &  {0.17 $\pm$ 0.02} &  {0.168 $\pm$ 0.005} &  {4.352 $\pm$ 0.004} & 1.197 & 28.309/32 \\ \noalign{\smallskip} \hline \noalign{\smallskip}
\end{tabular}
}}
\end{center}
\label{}
\end{table*}

\begin{table*}[p]
\caption{Values of fitted parameters with respect to the experimental
data on proton-spectra at different centralities of $Pb+Pb$ collisions at LHC energy $\sqrt{s_{NN}}=2.76$ TeV}
\begin{center}
\begin{tabular}{llllllll}
\hline \noalign{\smallskip}
Secondary & Centrality & {    $<N>$  } & {    $v_T(c)$  } & {    $T_{kin}$(GeV)  } & {$ c$   } & $q$ & $\chi^2/ndf$ \\
type &  &  &  &  &  &  &  \\ \hline \noalign{\smallskip}
$p$ & $0-5\%$ & 34$\pm$5 & 0.49$\pm$0.03 & 0.233$\pm$0.002 & 8.057$\pm$0.003 & 1.020 & 9.408/38 \\ 
 {} & $5-10\%$ & 28$\pm$4 & 0.49$\pm$0.02 & 0.233$\pm$0.003 & 7.528$\pm$0.003 & 1.028 & 7.924/38 \\ 
 {} & $10-20\%$ & 21.1$\pm$2.3 & 0.48$\pm$0.02 & 0.233$\pm$0.002 & 6.883$\pm$0.005 & 1.040 & 11.088/38 \\ 
 {} & $20-30\%$ & 14.7$\pm$1.1 & 0.47$\pm$0.02 & 0.234$\pm$0.004 & 6.972$\pm$0.004 & 1.044 & 10.789/38 \\ 
 {} & $30-40\%$ & 9.7$\pm$0.6 & 0.44$\pm$0.03 & 0.234$\pm$0.003 & 6.568$\pm$0.007 & 1.056 & 10.226/38 \\ 
 {} & $40-50\%$ & 6.1$\pm$0.3 & 0.41$\pm$0.03 & 0.234$\pm$0.003 & 6.564$\pm$0.006 & 1.066 & 9.330/38 \\ 
 {} & $50-60\%$ & 3.7$\pm$0.2 & 0.37$\pm$0.02 & 0.233$\pm$0.003 & 6.203$\pm$0.007 & 1.076 & 9.421/38 \\ 
 {} & $60-70\%$ & 1.9$\pm$0.3 & 0.33$\pm$0.03 & 0.233$\pm$0.002 & 6.311$\pm$0.003 & 1.099 & 9.223/38 \\ 
 {} & $70-80\%$ & 0.92$\pm$0.04 & 0.31$\pm$0.03 & 0.233$\pm$0.002 & 6.344$\pm$0.002 & 1.121 & 11.216/38 \\ 
 {} & $80-90\%$ & 0.392$\pm$0.004 & 0.29$\pm$0.02 & 0.232$\pm$0.005 & 6.669$\pm$0.004 & 1.146 & 26.723/38 \\  \hline \noalign{\smallskip}
$\bar{p}$ & $0-5\%$ & 33$\pm$3 & 0.49$\pm$0.02 & 0.237$\pm$0.003 & 8.656$\pm$0.004 & 1.024 & 10.686/38 \\ 
 {} & $5-10\%$ & 28$\pm$2 & 0.49$\pm$0.02 & 0.237$\pm$0.002 & 8.523$\pm$0.005 & 1.026 & 12.315/38 \\ 
 {} & $10-20\%$ & 21.2$\pm$1.8 & 0.49$\pm$0.02 & 0.239$\pm$0.005 & 8.086$\pm$0.004 & 1.034 & 11.195/38 \\ 
 {} & $20-30\%$ & 14.6$\pm$0.9 & 0.49$\pm$0.03 & 0.239$\pm$0.002 & 7.702$\pm$0.005 & 1.043 & 11.518/38 \\ 
 {} & $30-40\%$ & 9.8$\pm$0.3 & 0.46$\pm$0.01 & 0.239$\pm$0.003 & 7.329$\pm$0.008 & 1.050 & 10.967/38 \\ 
 {} & $40-50\%$ & 6.1$\pm$0.2 & 0.44$\pm$0.03 & 0.239$\pm$0.002 & 6.981$\pm$0.005 & 1.064 & 7.245/38 \\ 
 {} & $50-60\%$ & 3.7$\pm$0.2 & 0.39$\pm$0.03 & 0.239$\pm$0.002 & 6.634$\pm$0.006 & 1.073 & 10.088/38 \\ 
 {} & $60-70\%$ & 1.9$\pm$0.2 & 0.35$\pm$0.03 & 0.239$\pm$0.002 & 6.425$\pm$0.006 & 1.100 & 6.000/38 \\ 
 {} & $70-80\%$ & 0.92$\pm$0.02 & 0.32$\pm$0.02 & 0.239$\pm$0.003 & 6.342$\pm$0.003 & 1.124 & 5.479/38 \\ 
 {} & $80-90\%$ & 0.4$\pm$0.003 & 0.28$\pm$0.02 & 0.237$\pm$0.002 & 6.631$\pm$0.003 & 1.143 & 25.103/38 \\  
 \noalign{\smallskip} \hline \noalign{\smallskip}
\end{tabular}
\end{center}
\label{}
\end{table*}

\begin{table*}[h]
\caption{Values of $N_{part}$ and $N_{q-part}$ for different centralities of $Pb+Pb$ and $P+P$ collisions at 2.76 TeV}
\begin{center}
\begin{tabular}{lll}
\hline \noalign{\smallskip}
Centrality & $N_{part}$ & $N_{q-part}$ \\ \hline \noalign{\smallskip}
$0-5\%$ & 382.8 & 1014 \\ 
$5-10\%$ & 329.7 & 828 \\ 
$10-20\%$ & 260.5 & 625 \\ 
$20-30\%$ & 186.4 & 415 \\ 
$30-40\%$ & 128.9 & 267 \\ 
$40-50\%$ & 85 & 164 \\ 
$50-60\%$ & 52.8 & 91 \\ 
$60-70\%$ & 30 & 45 \\ 
$70-80\%$ & 15.8 & 18.9 \\ 
$80-90\%$ & 7.5 & 8 \\ 
$0-20\%$ & 308 & 773 \\ 
$20-40\%$ & 157 & 341 \\ 
$40-60\%$ & 69 & 127.5 \\ 
$60-80\%$ & 23 & 32 \\ \hline \noalign{\smallskip}
$pp$ & 2 & 3.05 \\ \noalign{\smallskip} \hline \noalign{\smallskip}
\end{tabular}
\end{center}
\label{}
\end{table*}

\clearpage

\newpage

\begin{figure*}[h]
\SetFigLayout{2}{2} \centering
\subfigure[]{\includegraphics[width=8cm]{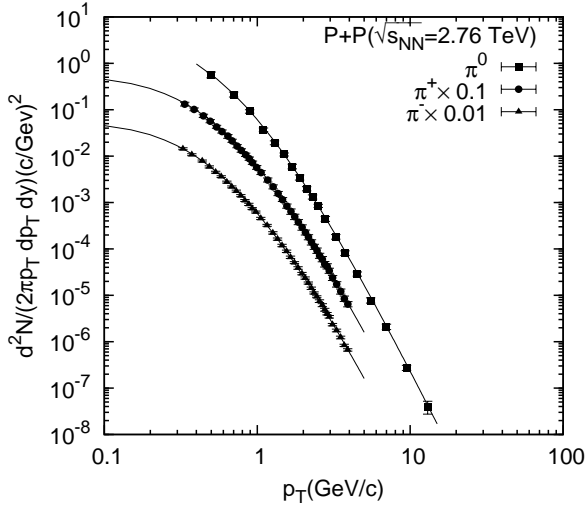}} \hfill
\subfigure[]{\includegraphics[width=8cm]{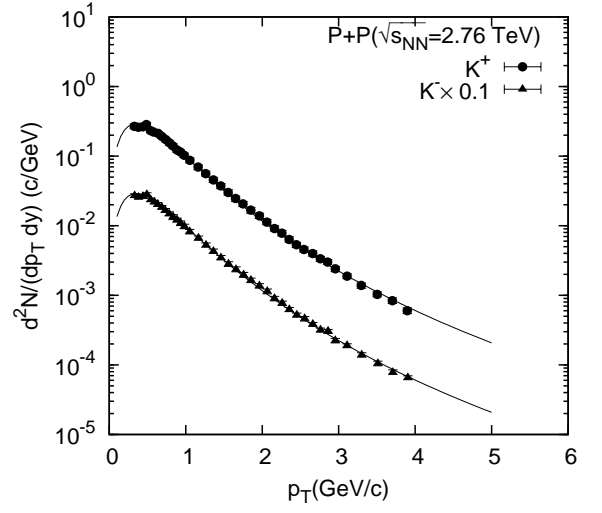}}\\
\subfigure[]{\includegraphics[width=8cm]{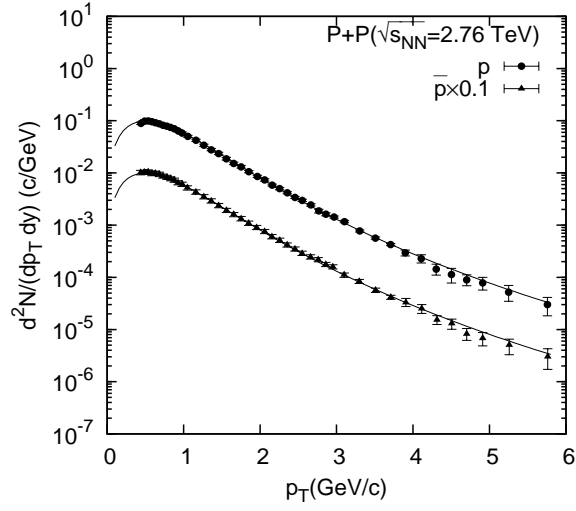}}
\caption{Plots of transverse momentum spectra of pions, kaons and protons
produced in $P+P$ collisions at $\sqrt{s_{NN}}=2.76$ TeV. The filled symbols represent the experimental data
points\cite{Peresunko1,Guerzoni1}. The solid curves provide the fits
on the basis of nonextensive approach(eqn.(13)).}
\end{figure*}

\begin{figure*}[h]
\SetFigLayout{2}{2} \centering
\subfigure[]{\includegraphics[width=8cm]{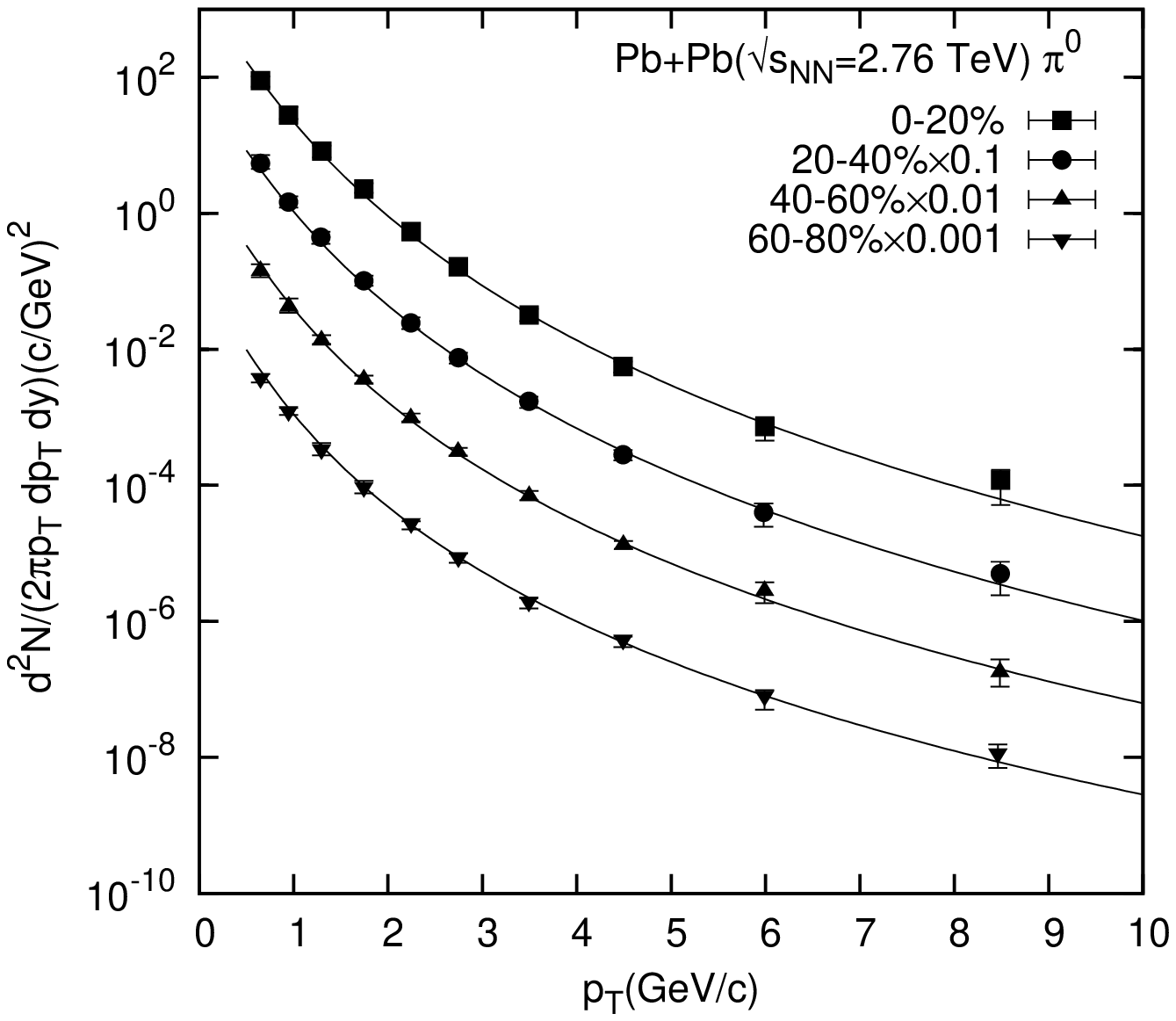}}\\
\subfigure[]{\includegraphics[width=8cm]{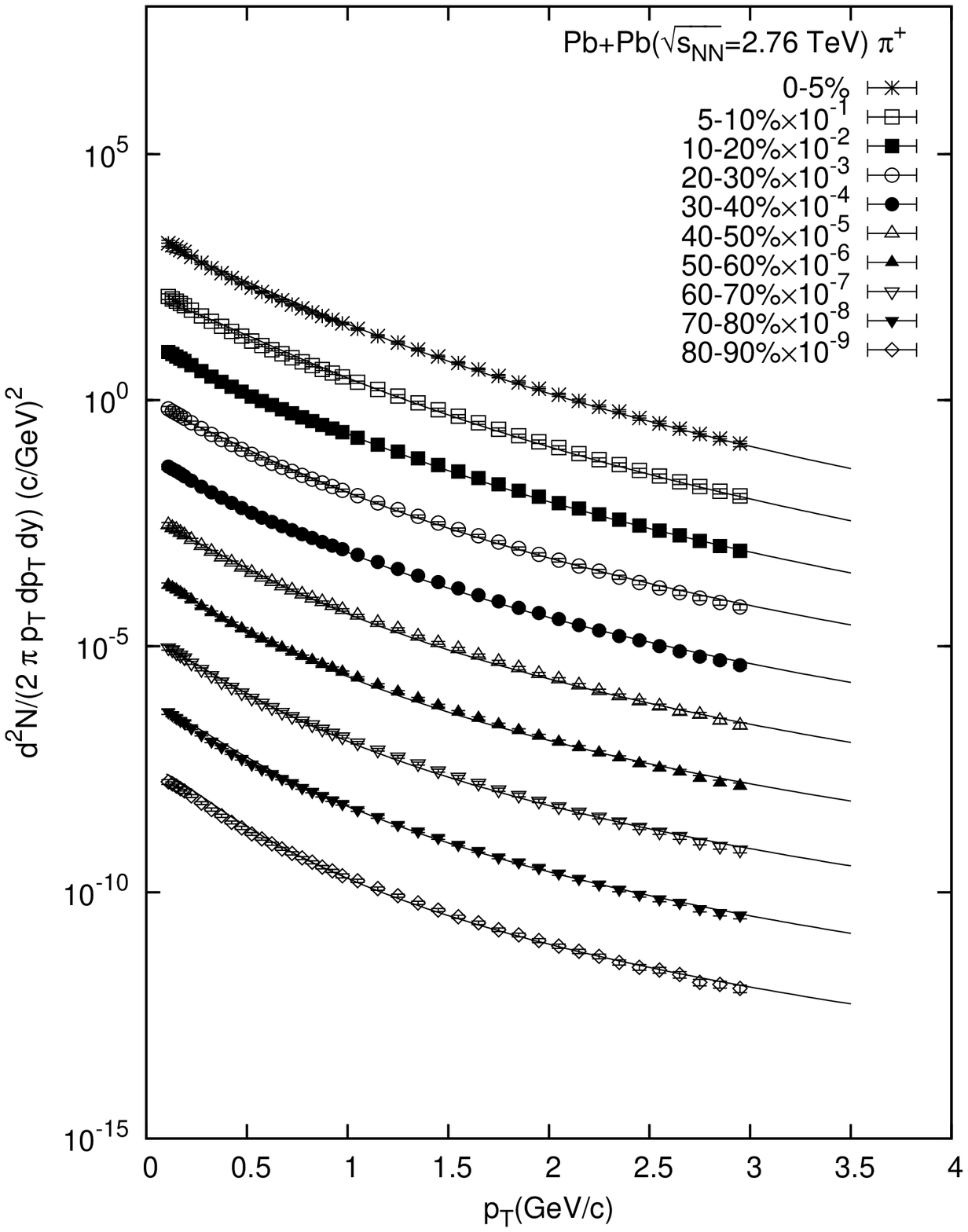}} 
\hfill
\subfigure[]{\includegraphics[width=8cm]{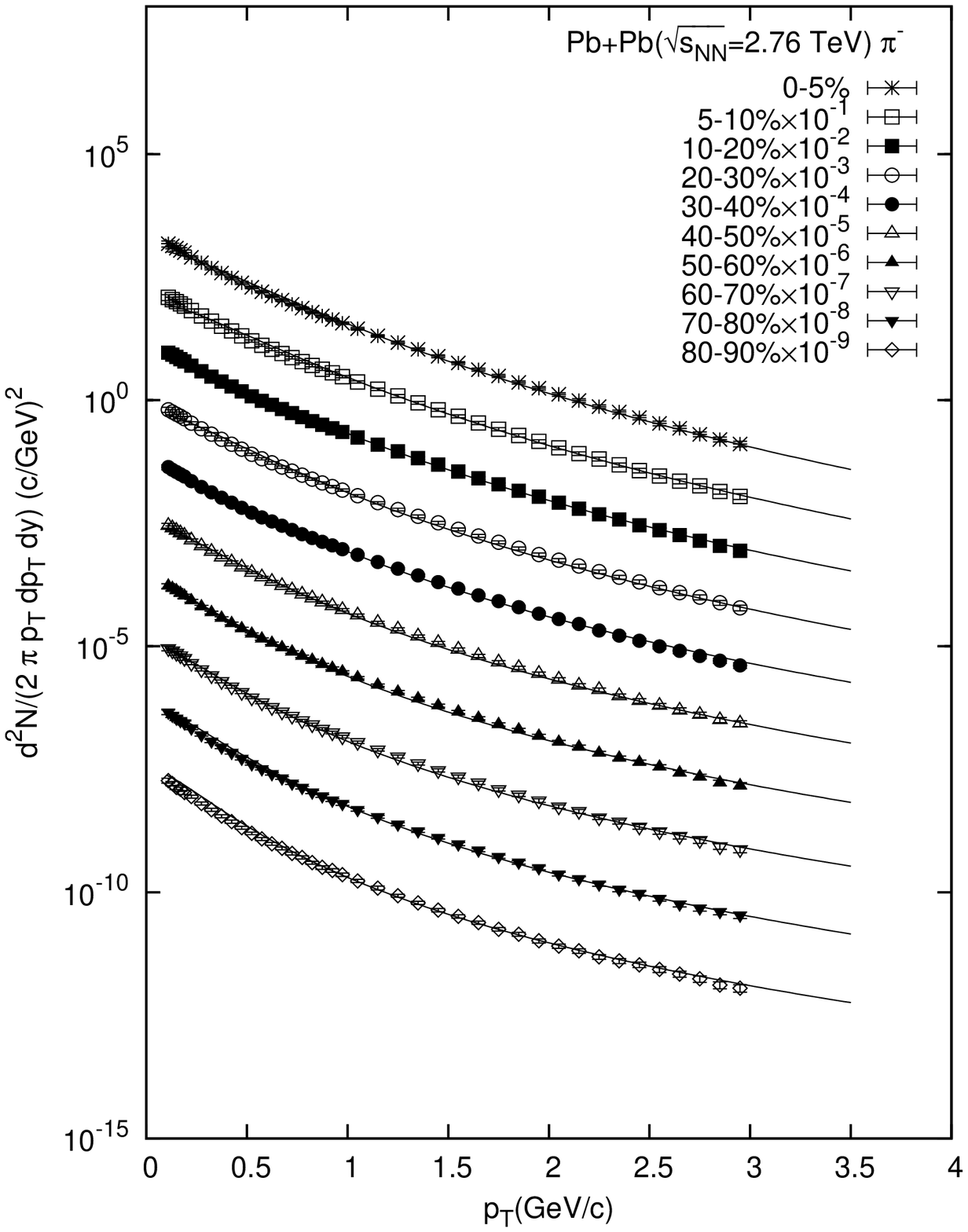}} 
\caption{Plots of Invariant yield of $\pi$-mesons
produced in different central $Pb+Pb$ collisions at $\sqrt{s_{NN}}=2.76$ TeV. The symbols represent the experimental data points\cite{Balbastre1,Abelev4} while the solid curves provide the fits
on the basis of nonextensive approach(eqn.(13,14,16)).}
\end{figure*}

\begin{figure*}[h]
\SetFigLayout{2}{2} \centering
\subfigure[]{\includegraphics[width=8cm]{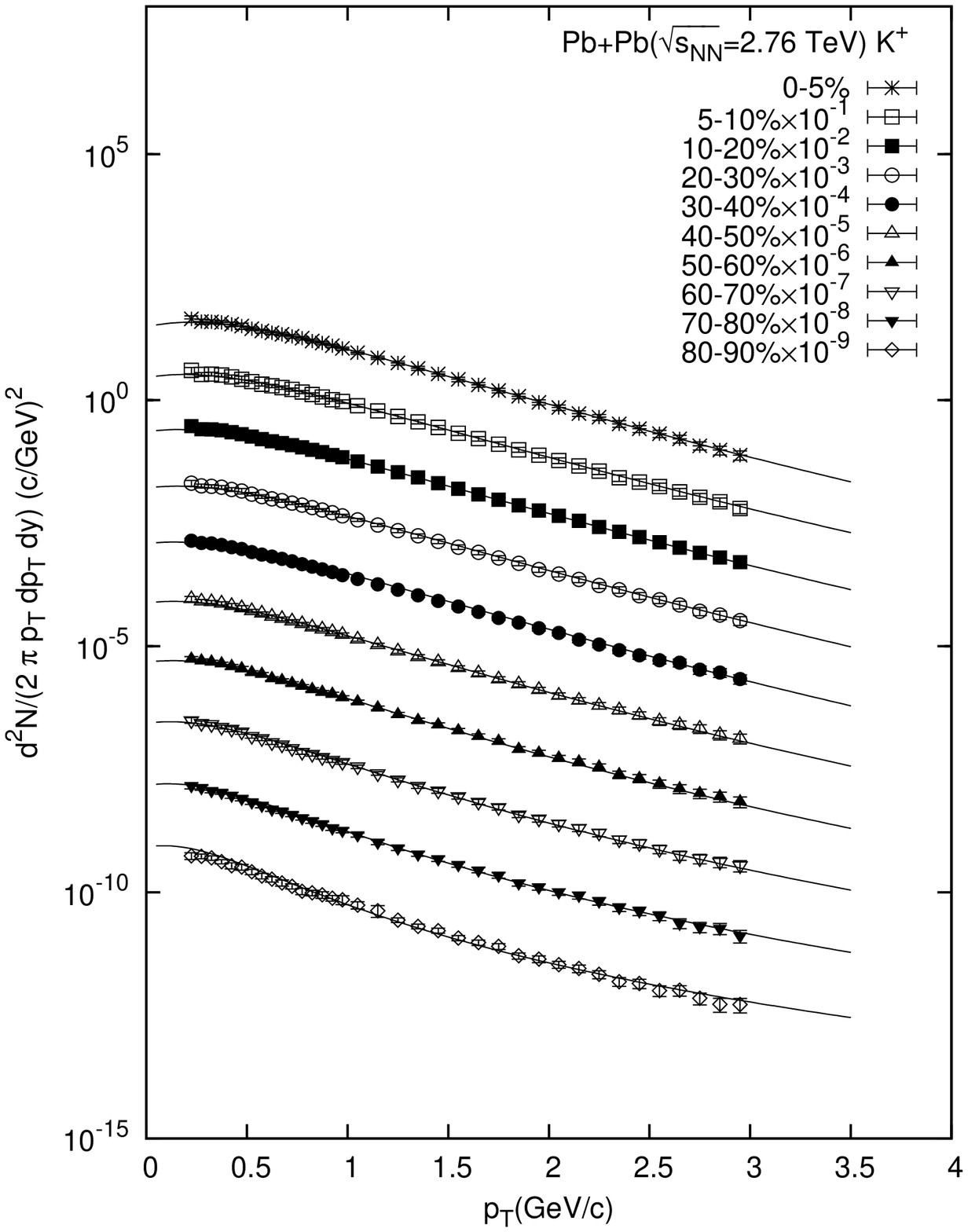}} 
\hfill
\subfigure[]{\includegraphics[width=8cm]{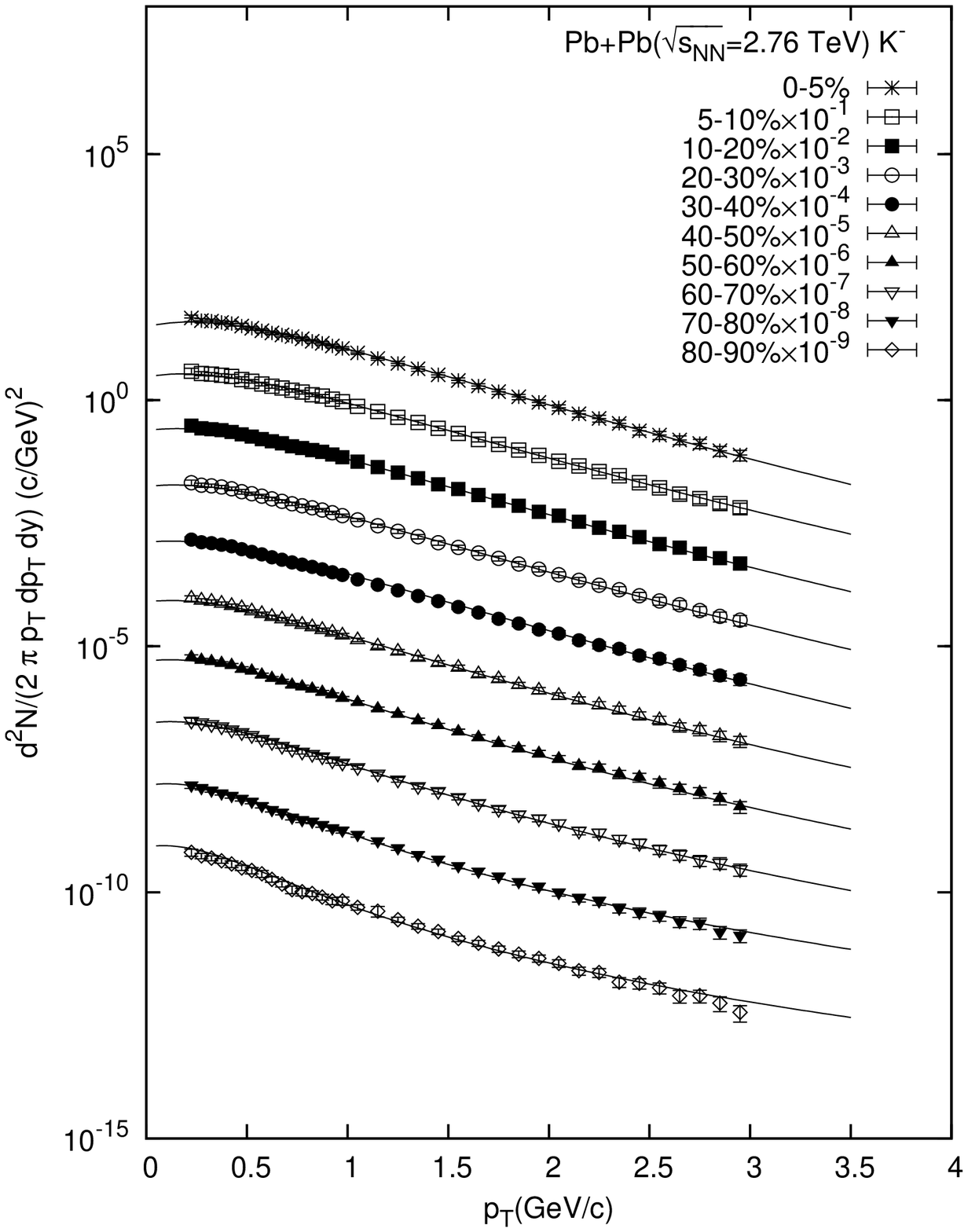}} 
\caption{Plots of Invariant yield of $K$-mesons
produced in different central $Pb+Pb$ collisions at $\sqrt{s_{NN}}=2.76$ TeV. The symbols represent the experimental data points\cite{Abelev4} while the solid curves provide the fits
on the basis of nonextensive approach(eqn.(13,14,16)).}
\end{figure*}

\begin{figure*}[h]
\SetFigLayout{2}{2} \centering
\subfigure[]{\includegraphics[width=8cm]{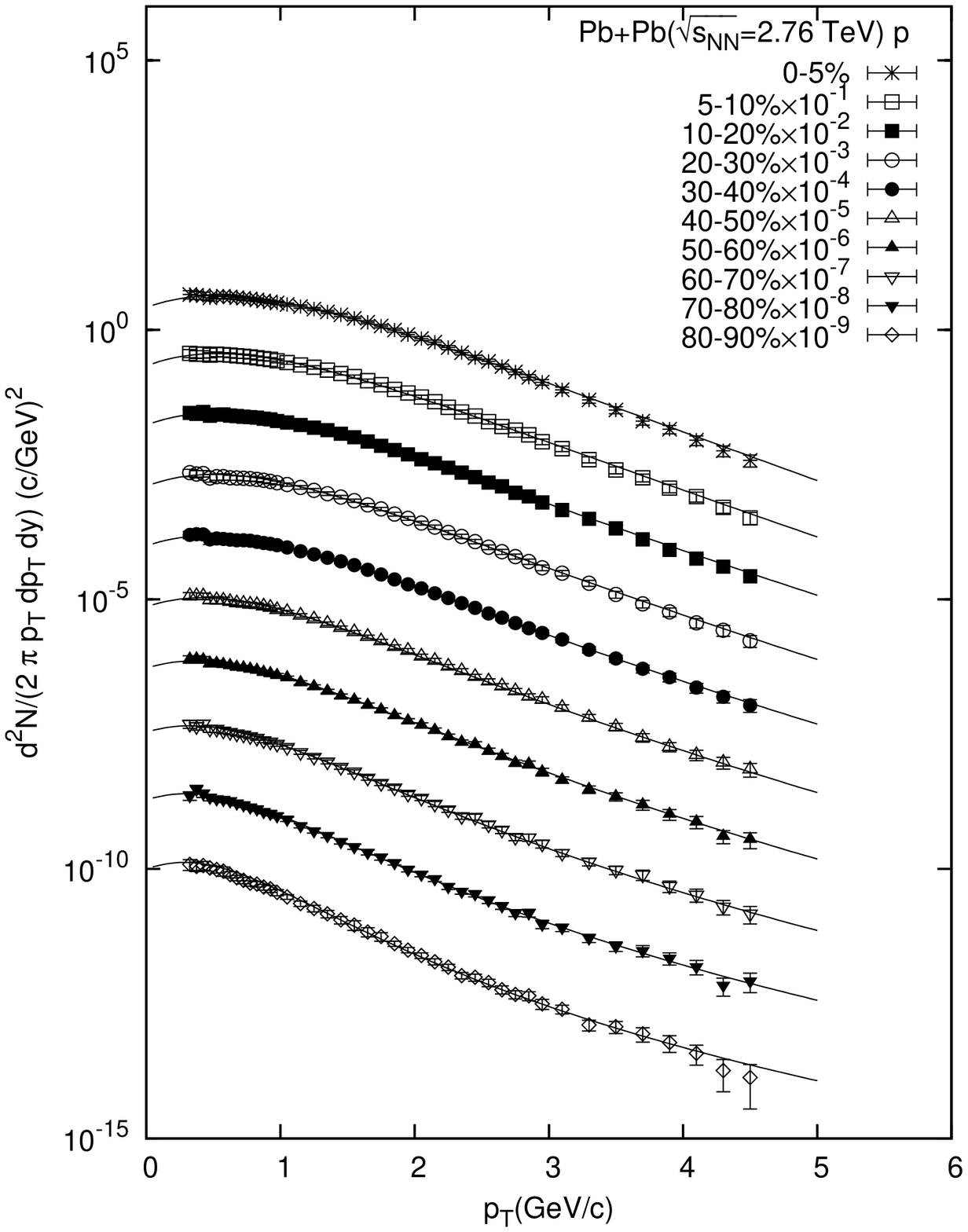}} 
\hfill
\subfigure[]{\includegraphics[width=8cm]{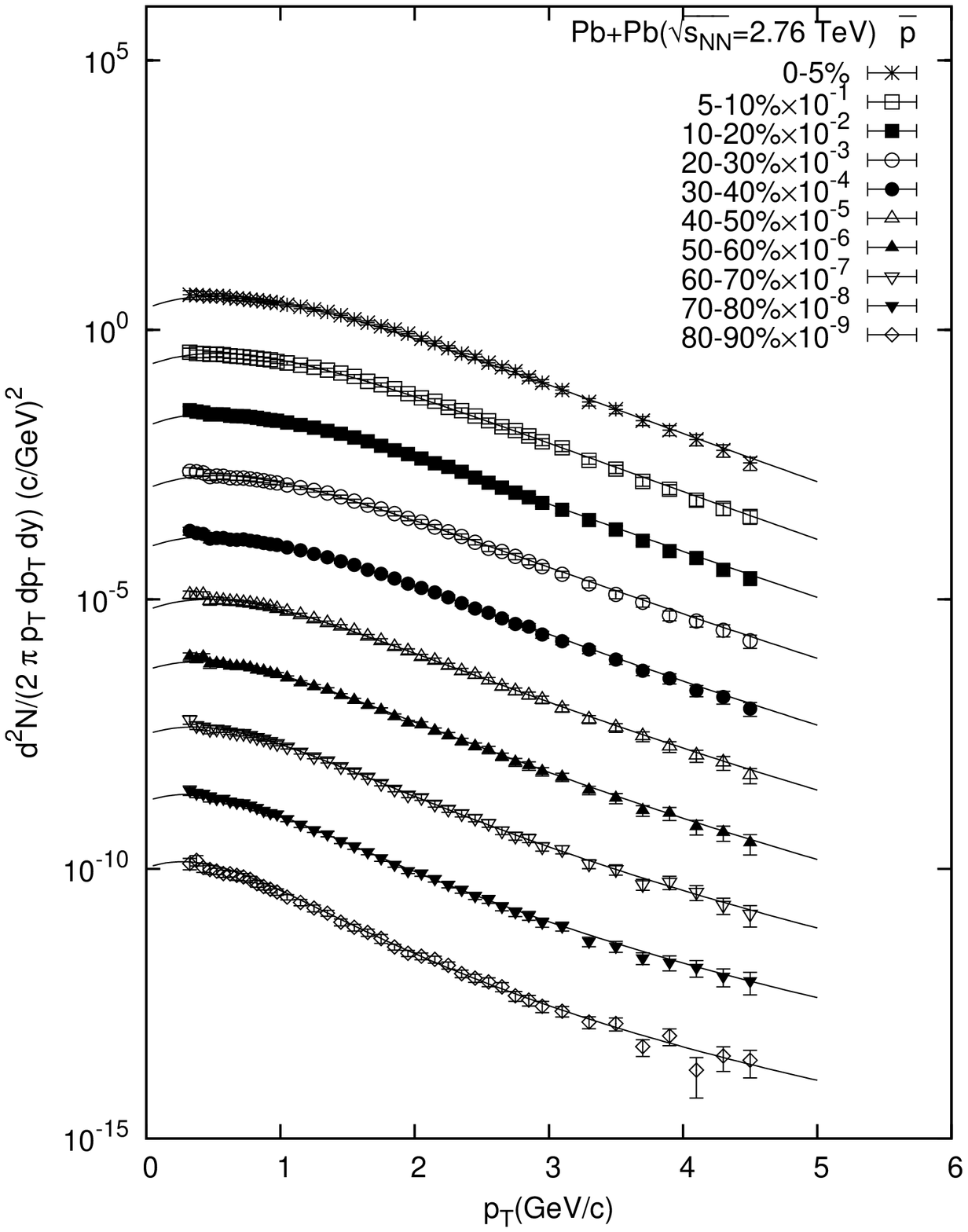}} 
\caption{Plots of Invariant yield of protons and anti-protons
produced in different central $Pb+Pb$ collisions at $\sqrt{s_{NN}}=2.76$ TeV. The symbols represent the experimental data points\cite{Abelev4} and the solid curves provide the fits
on the basis of nonextensive approach(eqn.(13,14,16)).}
\end{figure*}
\begin{figure*}[h]
\centering
\subfigure[]{\includegraphics[width=8cm]{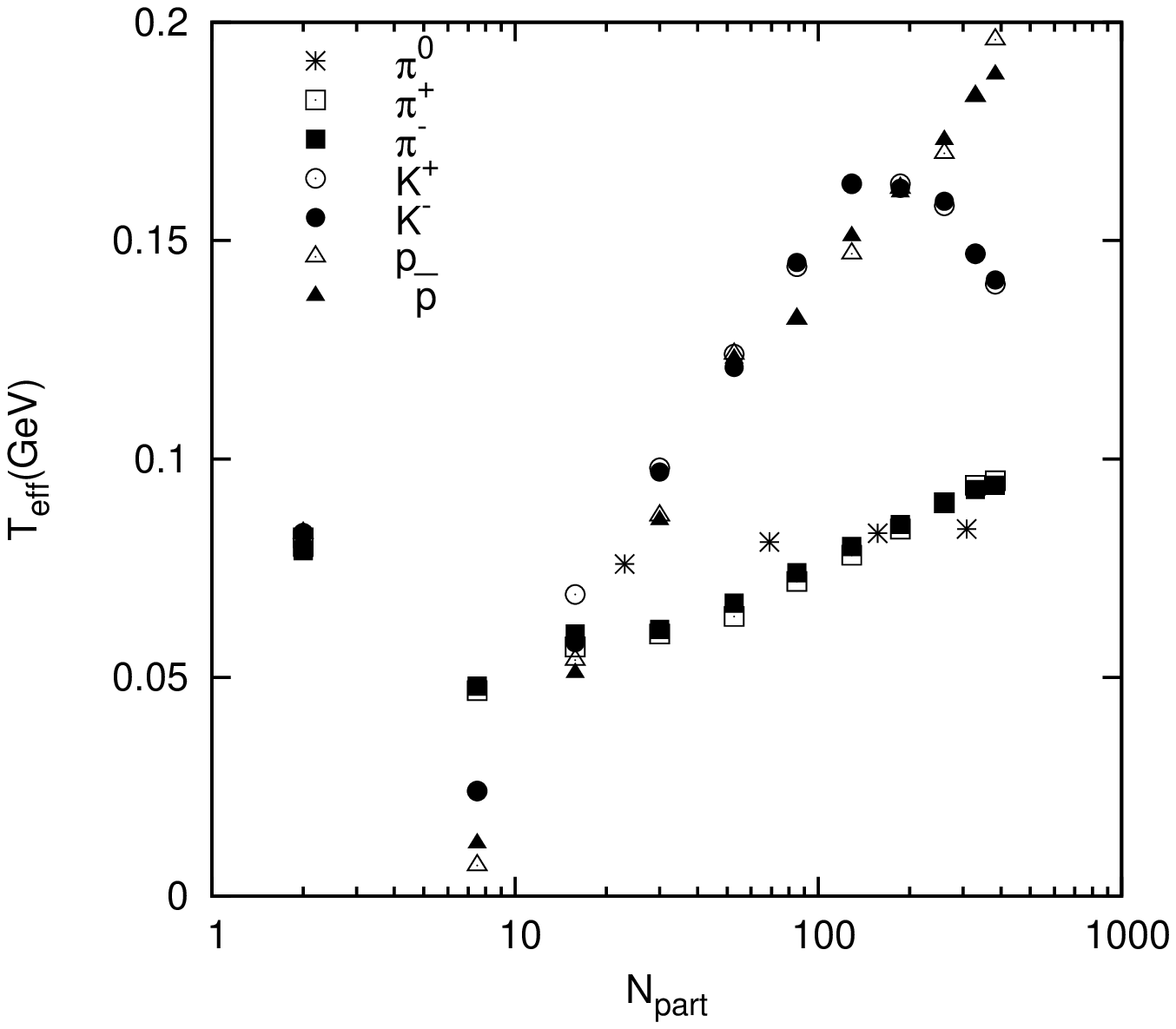}}\hfill
\subfigure[]{\includegraphics[width=8cm]{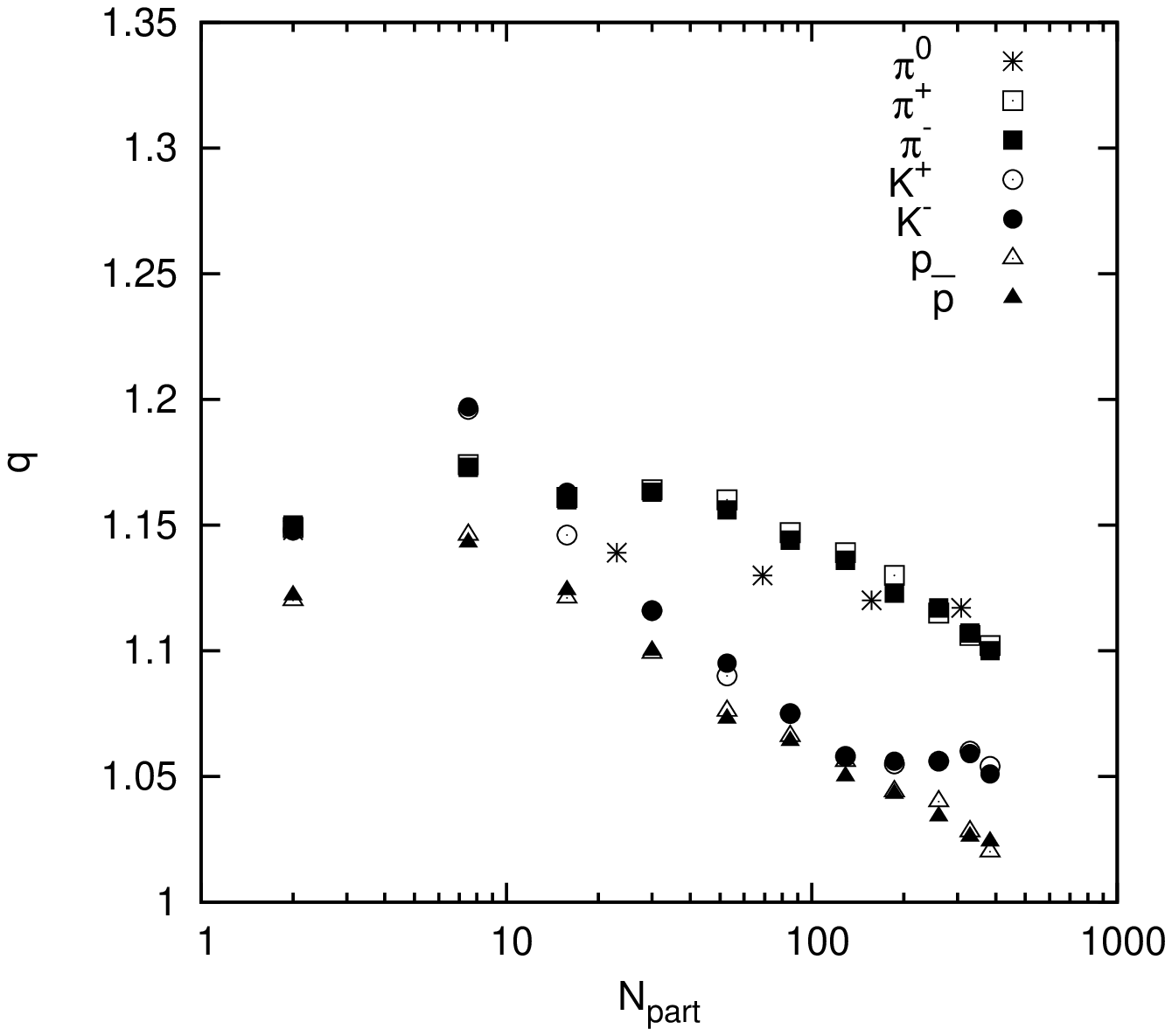}}\\
\subfigure[]{\includegraphics[width=8cm]{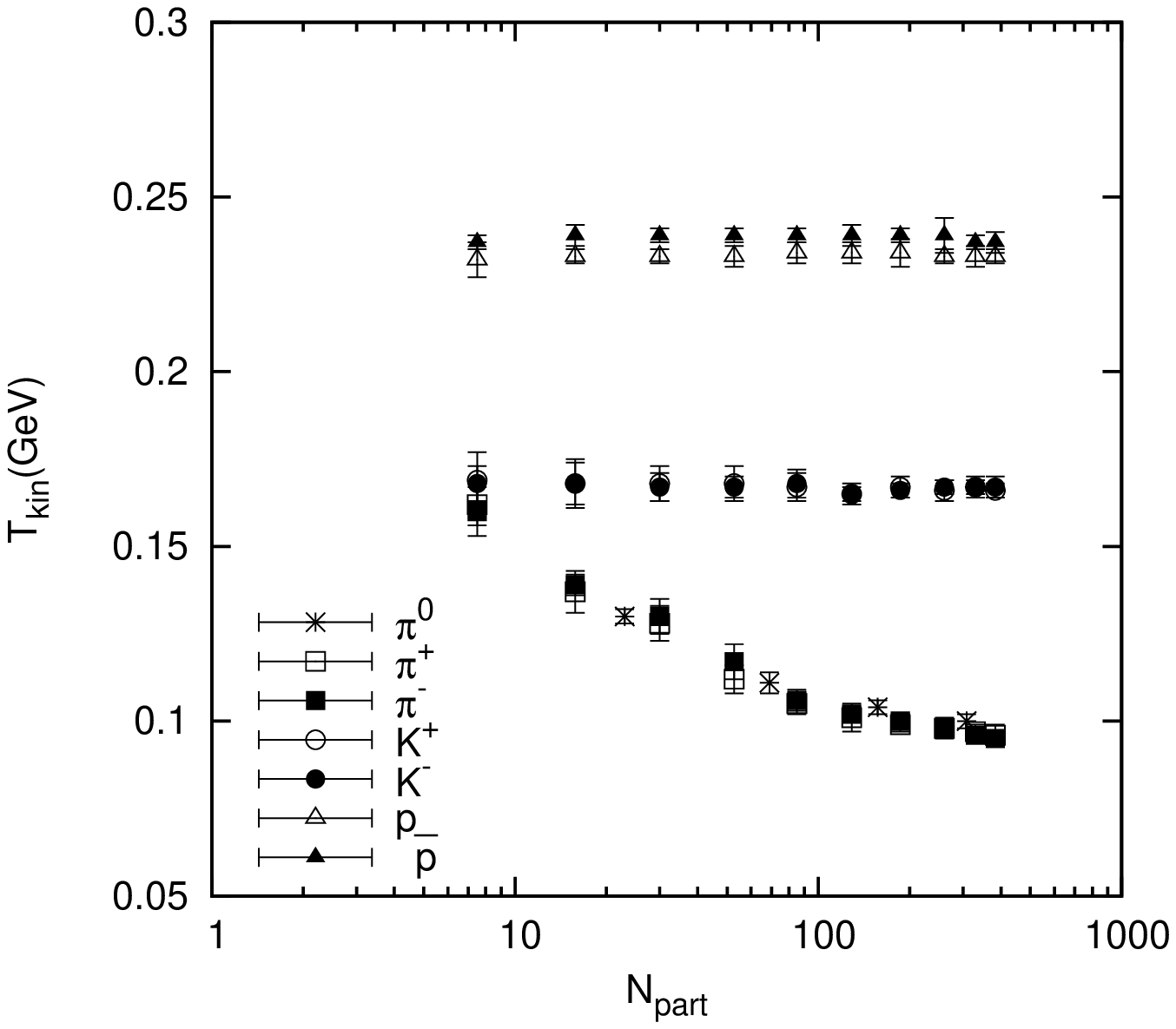}} \hfill
\subfigure[]{\includegraphics[width=8cm]{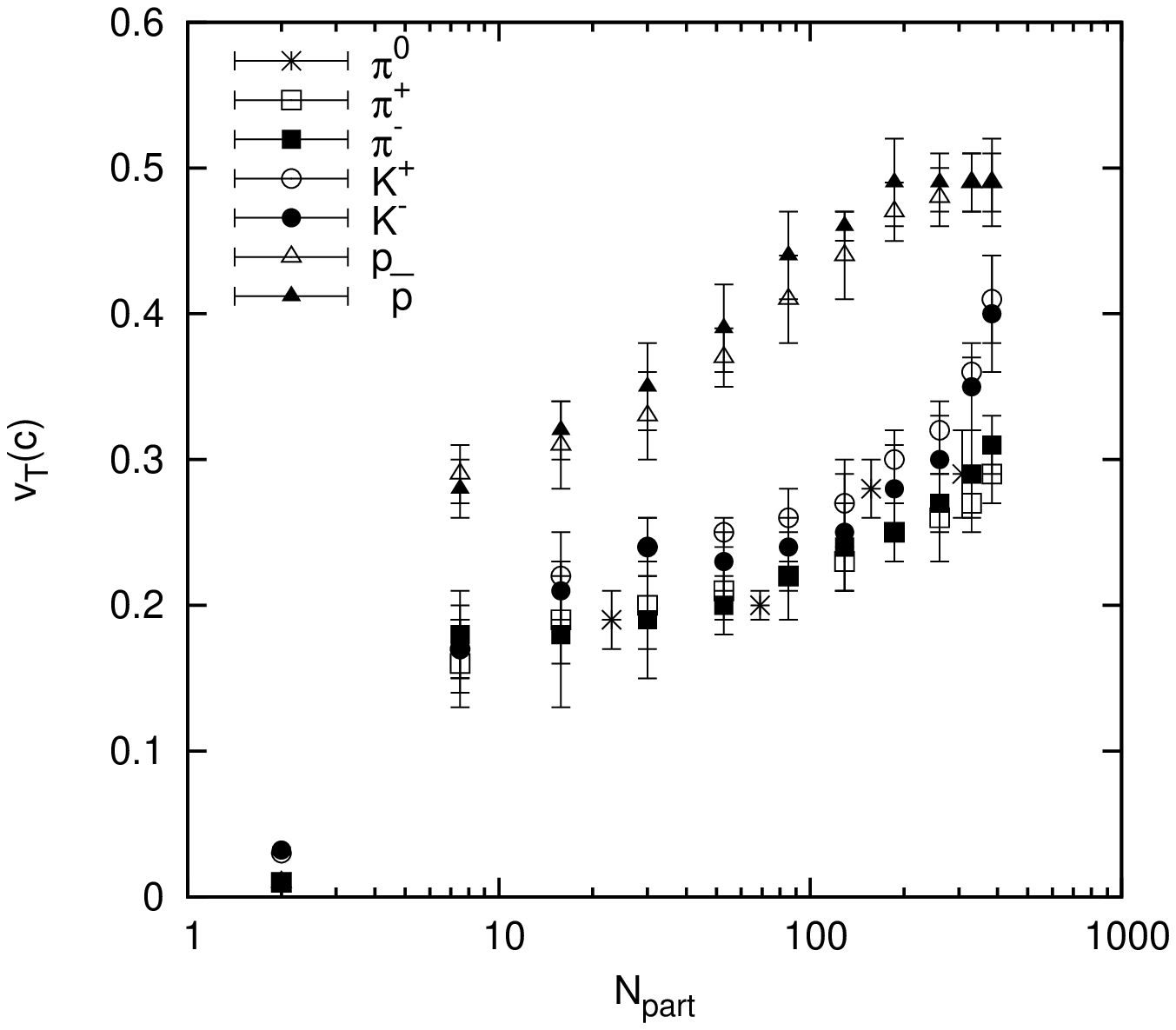}}
\caption{Plots of the effective temperature $T_{eff}$, the nonextensive parameter $q$, kinetic freeze-out temperature $T_{kin}$ and average transverse flow $v_T$,  obtained from the fits of different hadron-spectra for various centralities of $Pb+Pb$ collisions at $\sqrt{s_{NN}}=2.76$ TeV.}
\end{figure*}

\begin{figure*}[h]
\SetFigLayout{1}{2} \centering
\subfigure[]{\includegraphics[width=8cm]{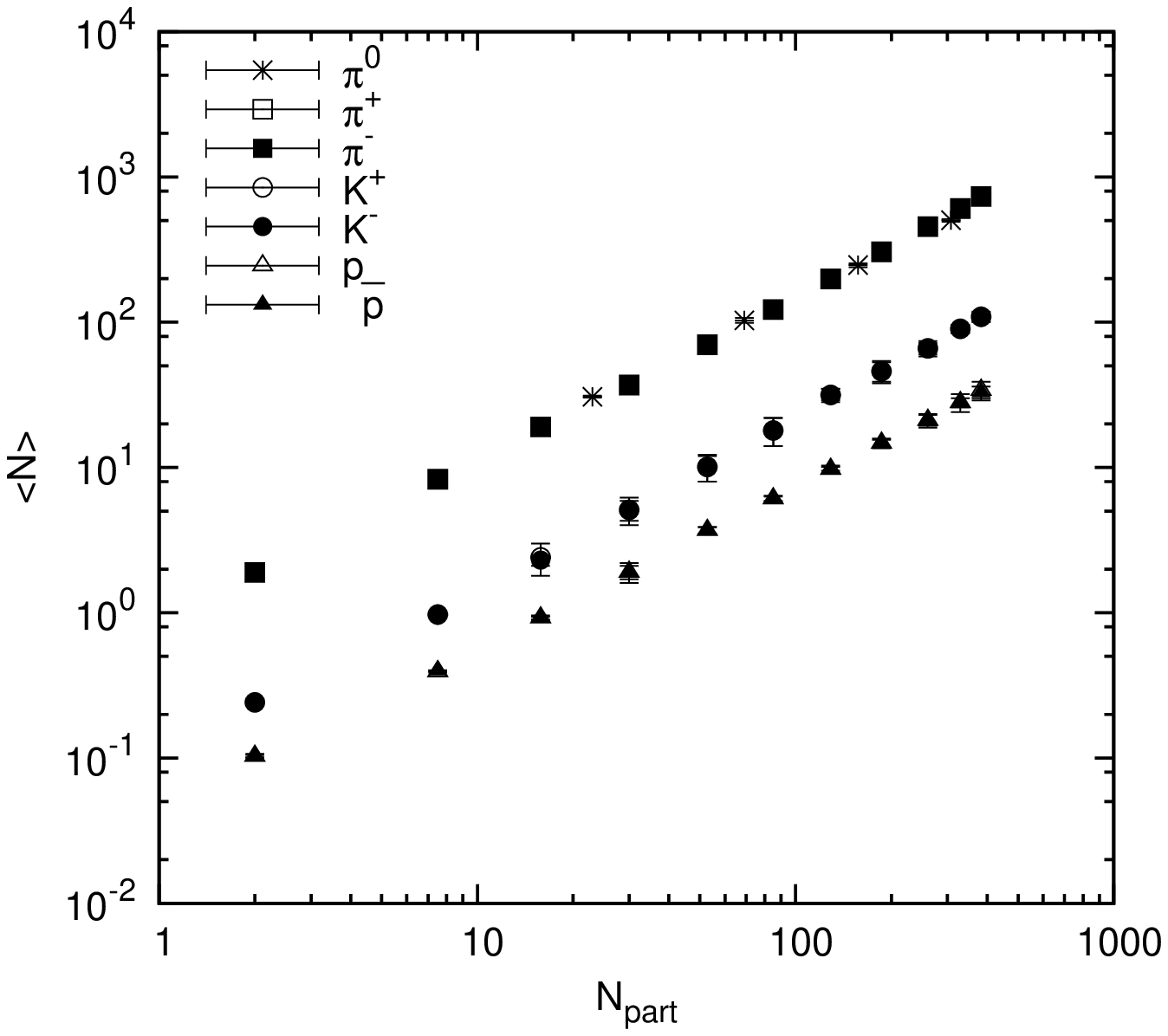}} \\
\subfigure[]{\includegraphics[width=8cm]{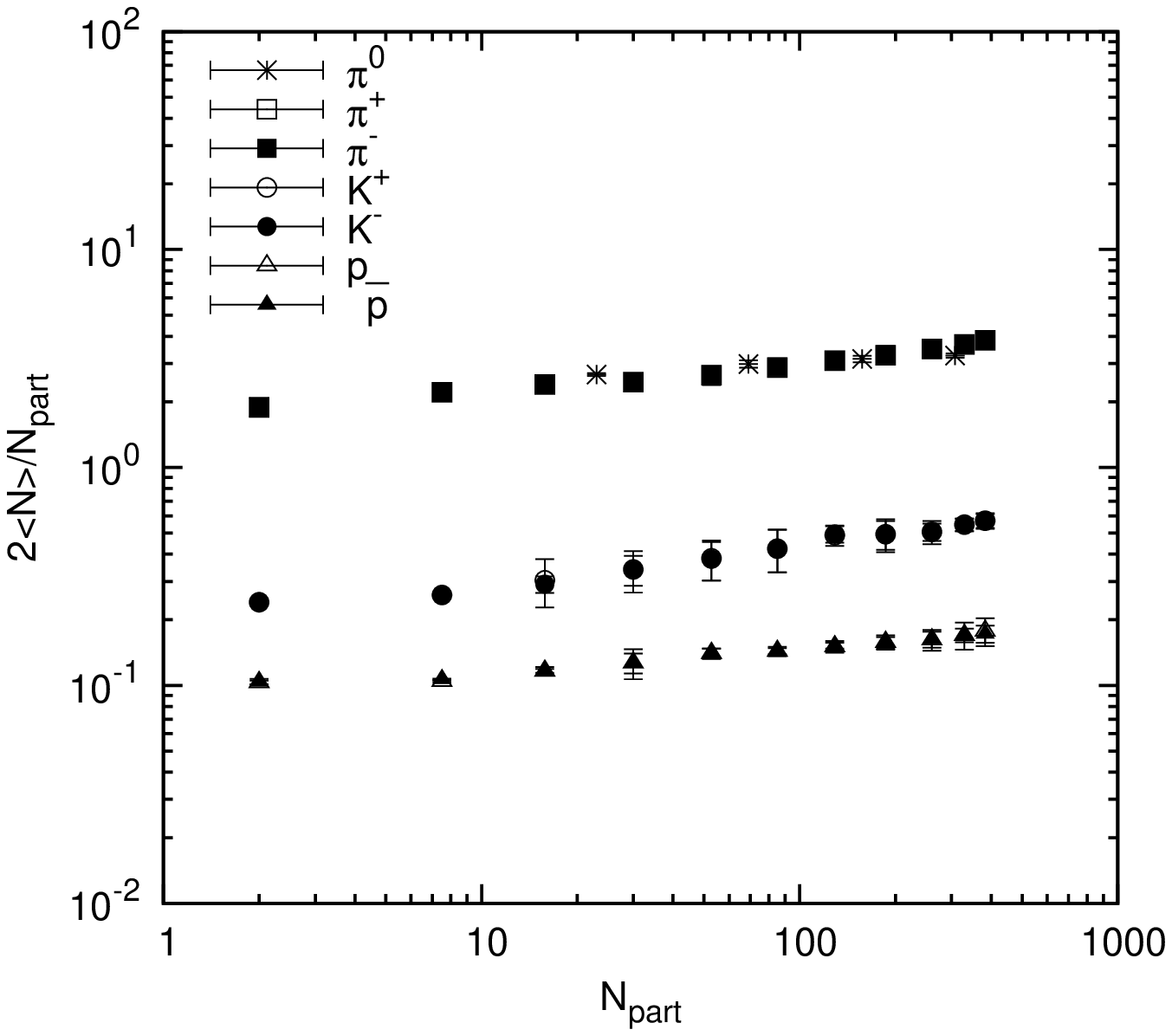}} 
\hfill
\subfigure[]{\includegraphics[width=8cm]{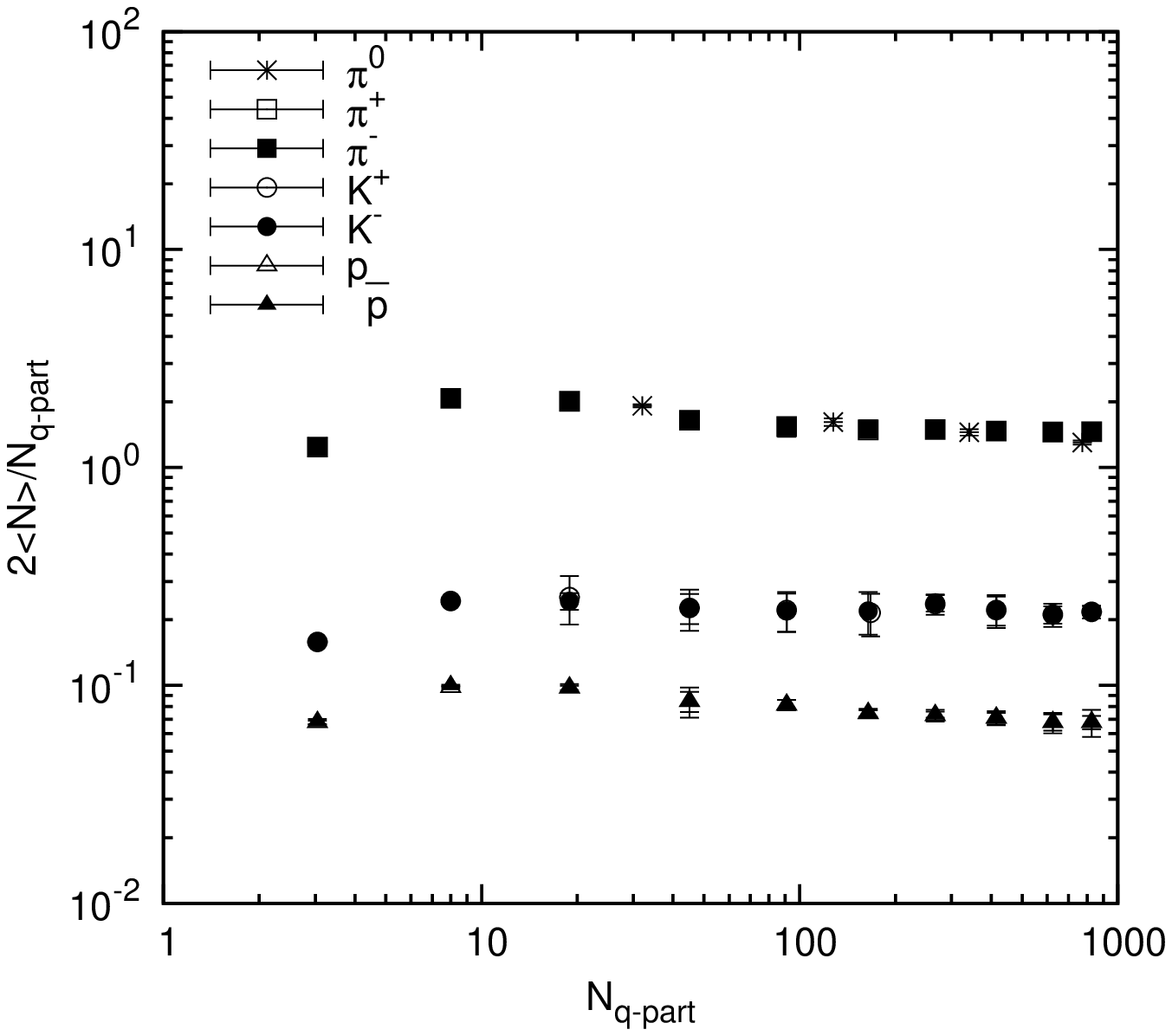}}
\caption{Plots of the average multiplicity for different hadrons produced in $P+P$ and $Pb+Pb$ collisions at LHC energy $\sqrt{s_{NN}}=2.76$ TeV; and the same while normalized by pair of participant nucleons and pair of participant quarks respectively.}
\end{figure*}


\newpage
\begin{thebibliography}{*}

\bibitem{De1} B. De, Eur. Phys. Jour. A {\bf 50}, 70 (2014).

\bibitem{Tsallis1} C. Tsallis, Jour. Stat. Phys. {\bf 52}, 479 (1988).

\bibitem{Tsallis2} C. Tsallis, {\it Nonextensive
Statistical Mechanics and It's Applications}, in {\it Lect. Notes Phys.}, Vol. {\bf{560}} (Springer,
2001) p. 3 . 

\bibitem{Tsallis3} C. Tsallis, Eur. Phys. Jour. A {\bf{40}}, 257 (2009).

\bibitem{Tsallis4} C. Tsallis, Braz. J. Phys. {\bf 39}, 337 (2009).

\bibitem{Tsallis5} C. Tsallis, Entropy {\bf 13}, 1765 (2011).

\bibitem{Prato1} D. Prato and C. Tsallis,  Phys. Rev. E {\bf 60}, 2398 (1999).

\bibitem{Beck1} C. Beck: Physica A {\bf 286}, 164 (2000). 

\bibitem{Beck2} C. Beck, Physica A {\bf 305}, 209 (2002).

\bibitem{Beck3} C. Beck, Eur. Phys. Jour A {\bf{40}}, 267 (2009).

\bibitem{Wilk1} G. Wilk and Z. Wlodarczyk,  Phys. Rev. Lett. {\bf{ 84}}, 2770 (2000).

\bibitem{Wilk1.1} G. Wilk and Z. Wlodarczyk,  Chaos Solitons Fractals {\bf 13}, 581 (2002). 

\bibitem{Wilk1.2} G. Wilk and Z. Wlodarczyk, Physica A {\bf 305}, 227 (2002).

\bibitem{Wilk1.3} G. Wilk and Z. Wlodarczyk,  AIP Conf. Proc. {\bf 965}, 76 (2007) arXiv:cond-mat/0708.2660. 

\bibitem{Wilk1.4} G. Wilk and Z. Wlodarczyk, Physica A {\bf 387}, 4809 (2008) arXiv:cond-mat/0711.3348. 

\bibitem{Wilk2} G. Wilk, Braz. J. Phys. {\bf 37}, 714 (2007) arXiv:hep-ph/0610292. 

\bibitem{Wilk3} G. Wilk and Z. Wlodarczyk, Phys. Rev. C {\bf 79}, 054903 (2009) arXiv:hep-ph/0902.3922.

\bibitem{Wilk4} G. Wilk and Z. Wlodarczyk,  Eur. Phys. Jour. A {\bf 40}, 299 (2009) arXiv:hep-ph/0810.2939.

\bibitem{Wilk5} G. Wilk and Z. Wlodarczyk,  Jour. Phys. G {\bf 38}, 065101 (2011). 

\bibitem{Osada1} T. Osada and G. Wilk, Phys. Rev. C {\bf 77}, 044903 (2008) arXiv:nucl-th/0710.1905.

\bibitem{Biro1} T. S. Biro and G. Purcsel, Phys. Rev. Lett {\bf 95}, 162302 (2005).

\bibitem{Biro2} T. S. Biro and K. Urmossy, Jour. Phys. G {\bf{36}}, 064044 (2009) arXiv:hep-ph/0812.2985. 

\bibitem{Biro3} T. S. Biro, G. Purcsel and K. Urmossy, Eur. Phys. Jour. A {\bf{40}}, 325 (2009). 

\bibitem{Biro3.1} T. S. Biro, K. Urmossy and Z. Schram, Jour. Phys. G {\bf 37}, 094027 (2010).

\bibitem{Biro4} T. S. Biro and E. Molnar, Eur. Phys. Jour. A {\bf{48}}, 172 (2012). 

\bibitem{Urmossy0.1} K. Urmossy, EPJ Web of Conferences {\bf 13}, 05003 (2011).

\bibitem{Biyajima1} M. Biyajima {\it et al}, Eur. Phys. Jour. C {\bf 40}, 243 (2005) arXiv:hep-ph/0403063.

\bibitem{Biyajima2} M. Biyajima {\it et al}, Eur. Phys. Jour. C {\bf 48}, 597 (2006) arXiv:hep-ph/0602120.

\bibitem{Alberico1} W. M. Alberico and A. Lavagno, Eur. Phys. Jour. A {\bf{40}}, 313 (2009) arXiv:nucl-th/0901.4952.

\bibitem{Lavagno1} A. Lavagno, P. Quarati and A. M. Scarfone, Braz. Jour. Phys. {\bf{39}}, 457 (2009). 

\bibitem{Lavagno2} A. Lavagno,  Phys. Lett. A {\bf{301}}, 13 (2002).

\bibitem{Kaniadakis} G. Kaniadakis, Eur. Phys. Jour. A {\bf{40}}, 275 (2009). 

\bibitem{Kodama} T. Kodama and T. Koide, Eur. Phys. Jour. A {\bf{A 40}}, 289 (2009).

\bibitem{De2} B. De et al,  Int. Jour. Mod. Phys. E {\bf 16}, 1687 (2007).

\bibitem{De3} B. De et al,  Int. Jour. Mod. Phys. A {\bf 25}, 1239 (2010).

\bibitem{Wibig1} T. Wibig, Jour. Phys. G {\bf 37}, 115009 (2010) arXiv:hep-ph/1005.5652. 

\bibitem{Wibig2} T. Wibig, Eur. Phys. Jour. C {\bf{74}}, 2966 (2014) arXiv:hep-ph/1304.0655v1.

\bibitem{Jiulin1} D. Jiulin, Chin. Phys. B {\bf 19}, 070501 (2010) arXiv:cond-mat/1012.2765; 

\bibitem{Jiulin2} G. Ran \& D. Jiulin, Physica A {\bf 391}, 2853 (2012) arXiv:cond-mat/1202.0638.

\bibitem{Urmossy1} K. Urmossy, arXiv:hep-ph/1212.0260v2

\bibitem{Urmossy2} P. Van, G. G. Barnafoldi, T. S. Biro and K. Urmossy, Jour. Phys. Conf. Ser. {\bf 394}, 012002 (2012) arXiv:stat-mech/1209.5963v1.

\bibitem{Deppman1} A. Deppman, Physica A {\bf 391}, 6380 (2012) arXiv: math-ph/1205.0455v2. 

\bibitem{Deppman2} I. Sena and A. Deppman, AIP Conf. Proc. {\bf 1520}, 172 (2013) arXiv: hep-ph/1208.2952v1. 

\bibitem{Deppman3} I. Sena and A. Deppman, Eur. Phys. Jour. A {\bf 49}, 17 (2013) arXiv: hep-ex/1209.2367v1. 

\bibitem{Deppman4} A. Deppman, Jour. Phys. G {\bf 41}, 055108 (2014) arXiv: hep-ph/1212.0379v2. 

\bibitem{Deppman5} L. Marques, E. Andrade-II and A. Deppman,  Phys. Rev. D {\bf 87}, 114022 (2013) arXiv: hep-ph/1210.1725v3. 

\bibitem{Cleymans1} J. Cleymans and D. Worku, Jour. Phys. G {\bf 39}, 025006 (2012).

\bibitem{Cleymans2} J. Cleymans and D. Worku, Eur. Phys. Jour. A {\bf 48}, 160 (2012).

\bibitem{Cleymans3} J. Cleymans, Phys. Lett. B {\bf{ 723}}, 351 (2013) arXiv:hep-ph/1309.7466.

\bibitem{Cleymans4} M. D. Azmi and J. Cleymans, arXiv: hep-ph/1310.0217.

\bibitem{Cleymans5} M. D. Azmi and J. Cleymans, Jour. Phys. G {\bf{41}}, 065001 (2014) arXiv: hep-ph/1401.4835. 

\bibitem{Rybczynski1} M. Rybczynski, Z. Wlodarczyk and G. Wilk, Jour. Phys. G {\bf 39}, 095004 (2012) arXiv: hep-ph/1203.6787v3.

\bibitem{Rybczynski2} M. Rybczynski, Z. Wlodarczyk and G. Wilk, Acta Phys. Pol. {\bf{B}} Proc. Suppl. {\bf{ 6}}, 507 (2013) arXiv: hep-ph/1212.1281.

\bibitem{Ristea1} O. Ristea {\it et al}, Jour. of Phys. Conf. Ser. {\bf 420}, 012041 (2013).

\bibitem{Khandai1} P. K. Khandai {\it et al}, Jour. of Phys. G {\bf 41}, 025105 (2014).

\bibitem{Lee1} K. S. Lee, U. Heinz and E. Schneddermann, Z. Phys. C {\bf 48}, 525 (1990).

\bibitem{Blaizot1} J. P. Blaizot and J. Y. Ollitrault, Adv. Ser. Direct. High Energy Phys. {\bf 6}, 393 (1990).

\bibitem{Bearden1} I. G. Bearden {\it et al}, Phys. Rev. Lett. {\bf 78}, 2080 (1997).

\bibitem{Abelev4} ALICE Collaboration (B. Abelev {\it{et al}}), Phys. Rev. C {\bf 88}, 044910 (2013).

\bibitem{Abelev2}  ALICE Collaboration (B. Abelev {\it{et al}}), Jour. High. Ener. Phys. {\bf 09}, 112 (2012) arXiv:nucl-ex/1203.2160.

\bibitem{Aamodt1} ALICE Collaboration (K. Aamodt {\it et al}), Phys. Rev. Lett. {\bf 106}, 032301 (2011).

\bibitem{Abelev3} ALICE Collaboration (B. Abelev {\it{et al}}), Phys. Rev. C {\bf 88}, 044909 (2013).

\bibitem{Alver1} B. Alver {\it et al.}, arXiv:nucl-ex/0805.4411.

\bibitem{Voloshin1} S. Eremin \& S. Voloshin,  Phys. Rev. C {\bf 67}, 064905 (2003).

\bibitem{Netrakanti1} P. K. Netrakanti \& B. Mohanty: Phys. Rev. C {\bf 70}, 027901 (2004). 

\bibitem{De4} B. De \& S. Bhattacharyya: Phys. Rev. C {\bf 71}, 024903 (2005).

\bibitem{Peresunko1} ALICE Collaboration (D. Peresunko), Nucl. Phys. A {\bf 904-905}, 755c (2013) arXiv:nucl-ex/1210.5749

\bibitem{Guerzoni1} ALICE Collaboration (B. Guerzoni), {\it pp spectra at 2.76 TeV:Summary},  http://agenda.infn.it/materialDisplay.py?contribId$=0$ \&materialId$=$slides\&confId$=6125$.

\bibitem{Balbastre1}  ALICE Collaboration (G. Conesa Balbastre): Jour. Phys. G {\bf{38}}, 124117 (2011) arXiv:hep-ex/1109.4929

\end{thebibliography}
\end{document}